\documentclass[preprint,12pt]{elsarticle}
\usepackage{graphics,color}
\usepackage{amssymb}
\usepackage{amsthm}
\usepackage{amsmath,natbib}%,fleqn}
\usepackage{hyperref}
\newcommand{\bw}{\mathbf{w}}
\newcommand{\bx}{\mathbf{x}}
\newcommand{\bM}{\mathbf{M}}

\newcommand{\bT}{\mathbf{T}}
\newcommand{\bz}{\mathbf{z}}

\newcommand{\bsr}{\boldsymbol{r}}
\newcommand{\bst}{\boldsymbol{t}}

\newcommand{\bsw}{\boldsymbol{w}}

\newcommand{\bsX}{\boldsymbol{X}}

\newcommand{\bsbeta}{\boldsymbol{\beta}}
\newcommand{\bsgamma}{\boldsymbol{\gamma}}
\newcommand{\bstheta}{\boldsymbol{\theta}}

\newcommand{\bspsi}{\boldsymbol{\psi}}

\newcommand{\bsepsilon}{\boldsymbol{\epsilon}}

\newcommand{\IR}{I\!\!R}
\journal{Neurocomputing}
\begin{document}
\begin{frontmatter}
\title{A hidden process regression model for functional data description. Application to curve discrimination}
\author[inrets,utc]{Faicel Chamroukhi\corref{cor1}}
\cortext[cor1]{Corresponding author:\\Faicel Chamroukhi\\ INRETS, 2 Rue de la Butte Verte,\\ 93166 Noisy-le-Grand Cedex, France\\Tel: +33(1) 45 92 56 46\\Fax: +33(1) 45 92 55 01}
\ead{faicel.chamroukhi@inrets.fr}
\author[inrets]{Allou Sam\'{e}}
\author[utc]{G\'erard Govaert}
\author[inrets]{Patrice Aknin}
\address[inrets]{French National Institute for Transport and Safety Research (INRETS)\\
Laboratory of New Technologies (LTN)\\2 Rue de la Butte Verte, \\ 93166 Noisy-le-Grand Cedex (France)\fnref{label1} \\ \ }
\address[utc]{Compi\`{e}gne University of Technology \\HEUDIASYC Laboratory, UMR CNRS 6599 \\
BP 20529, 60205 Compi\`{e}gne Cedex (France)}

\begin{abstract}
A new approach for functional data description is proposed in this paper. It consists of a regression model with a discrete hidden logistic process which is adapted for modeling curves with abrupt or smooth regime changes. The model parameters are estimated in a maximum likelihood framework through a dedicated Expectation Maximization (EM) algorithm.  From the proposed generative model, a curve discrimination rule is derived using the Maximum A Posteriori rule. The proposed model is evaluated using simulated curves and real world curves acquired during railway switch operations, by performing comparisons with the piecewise regression approach in terms of curve modeling and classification.

\end{abstract}


\end{frontmatter}

\section{Introduction}
\label{sec: introduction}

\begin{figure*}[!h]
\centering
\begin{tabular}{cc}
\includegraphics[width=6.5cm]{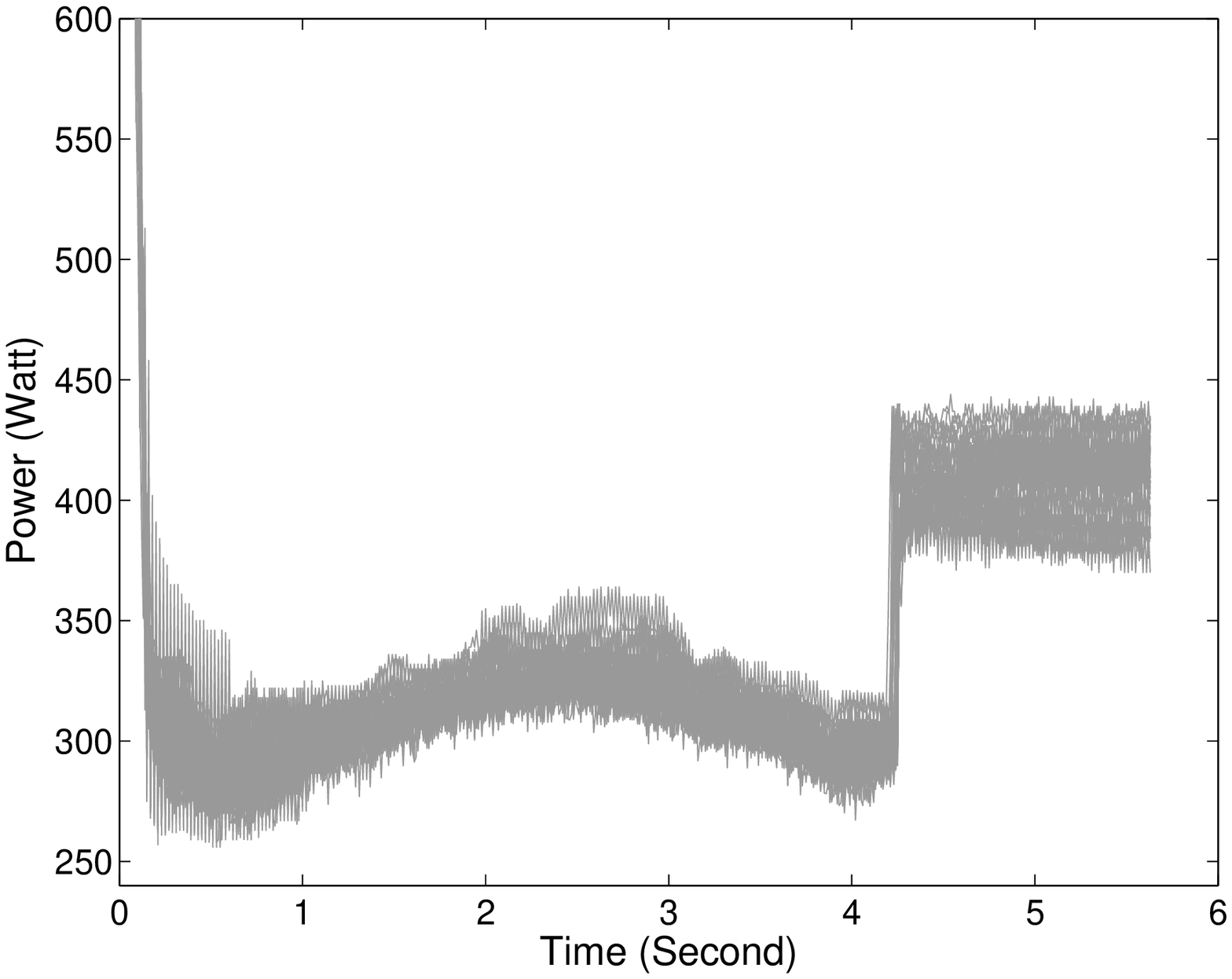}&
\includegraphics[width=6.5cm]{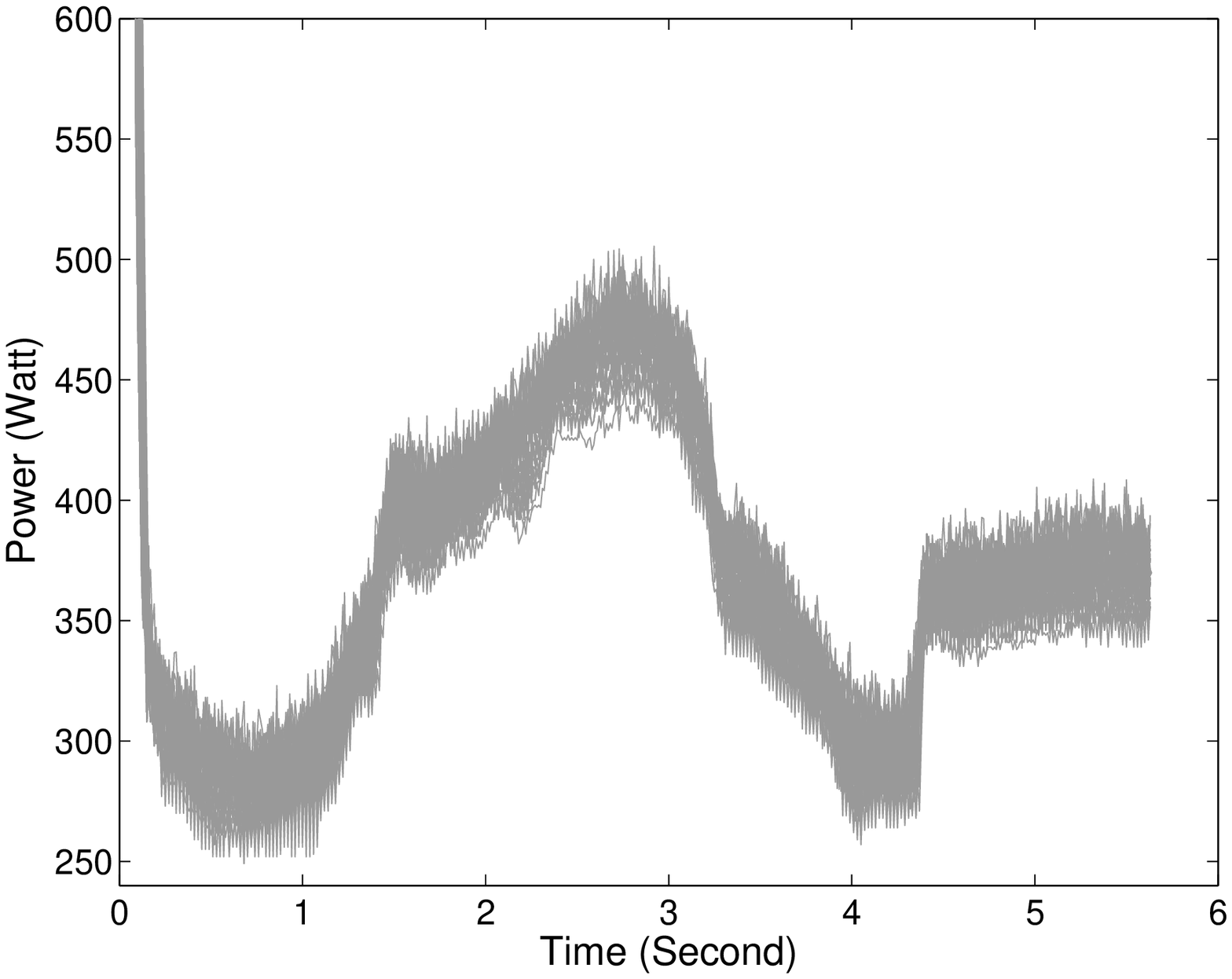}\\
\small (a)&\small (b)
\end{tabular}
\caption{Examples of curves of electrical power consumed during
various switch operations: 35 curves correspond to operations
without defect (a) and 45 curves correspond to operation with
critical defect (b).} \label{signal_intro}
\end{figure*}

Curve valued or functional datasets are increasingly available in
science, engineering and economics.   The work presented in this paper relates to the diagnosis of the French railway switches (or points) which enables trains to be guided from one track to another at a railway
junction. The switch is controlled by an electrical motor and the
considered curves are the condition measurements acquired during
switch operations. Each curve represents the electrical power
consumed during a switch operation (see Fig. \ref{signal_intro}). 

 To achieve the diagnosis task, the acquired curves have to be
accurately summarized. Summarizing  these curves can be performed by finding a simplified representation of each class of curves using an adapted model. 

The switch operations curves can be seen as functional data presenting non-linearities and different changes in regime due to the mechanical motions involved
in a switch operation (see fig. \ref{signal_intro}). In this
context, basic polynomial regression models can not be used to find an accurate description of such data. An alternative approach consists in using splines to approximate each set
of curves \cite{garetjamesANDtrevorhastieJRSS2001,garetjamesJRSS2002} but this requires the setting of knots. Another
approach, that allows for fitting several (polynomial) models to the curves for different time ranges,  consists in the piecewise polynomial regression model  used in \cite{McGee,brailovsky,ferrari1, HugueneyEtAl:ESANN2009} for curve approximation and segmentation. A related segmentation method  is the segmentation-clustering approach of Picard et al. \cite{picardetal2007} applied to array CGH data. One can also distinguish the recently proposed algorithm of Hugueney et al.  \cite{HugueneyEtAl:ESANN2009}  for curves clustering and segmentation using piecewise regression.

%In a functional data clustering context, one can distinguish the generative methods developed by Gaffney \& Smyth \cite{gaffneyANDsmythNIPS2004} and Liu \& Yang \cite{liuANDyangFunctionalDataClustering} and the piecewise regression approach for curves   clustering and segmentation recently proposed by  Hugueney et al.  \cite{HugueneyEtAl:ESANN2009}.  

Let's recall that piecewise polynomial regression is a
representation and segmentation method that partitions curves into
K segments (or regimes), each segment being characterized by its
mean polynomial curve and its variance. The parameters estimation is
performed using Fisher's algorithm \cite{fisher} which globally optimizes an
additive cost function \cite{yveslechevalier90} using a
dynamic programming procedure \cite{bellman}. The piecewise regression model is
more adapted for modeling curves presenting abrupt changes and is
less efficient for curves including regimes with smooth transitions.
Moreover, the dynamic programming procedure is computationally
expensive, especially for large samples.

In this paper, a generative model is explored to give a synthetic
representation of a set of curves presenting % both abrupt and smooth
changes in regime. The basic idea of the proposed model is to fit a
specific regression model incorporating a discrete hidden process
allowing for abrupt or smooth transitions between different
polynomial regression models. This approach is an extension, to a set of curves, of the
works presented in \cite{chamroukhiESANN2009, chamroukhi et al. NN2009}. It is related to the
switching regression model introduced in \cite{quandt} and is very linked to
the Mixture of Experts (ME) model \cite{jordanHME, waterhouse} by the using of a
time-dependent logistic transition function. 

In addition to providing a simplified representation of functional
data, the proposed model can be used for curve discrimination
through the Maximum a Posteriori (MAP) rule. A related method is  the Functional Linear Discriminant
Analysis (FLDA) \cite{garetjamesANDtrevorhastieJRSS2001}, where a cubic spline is used for curves
approximation. More details about the Functional Data Analysis (FDA) framework can be found in \cite{ramsayandsilvermanFDA2005, ramsayandsilvermanAppliedFDA2002}. Other works for curve classification include
neural network approaches \cite{RossiConanGuez05NeuralNetworks} and kernel-based learning methods \cite{RossiAndVillaSVM_functional data_Neurocomputing_2006}.

This paper is organized as follows. Section 2 provides an account of
the piecewise polynomial regression model in the context of modeling
a set of curves and the used parameter estimation technique based on
dynamic programming. Section 3 introduces the proposed model for
functional data representation and provides details of the
parameters estimation by means of a dedicated EM algorithm. The
curves approximation with the proposed approach is then presented
and a curve classification scheme using the MAP rule is used. Section 4
deals with the experimental study carried out on simulated curves and  real switch operation curves to asses the proposed approach.

\section{Modeling a set of curves by the piecewise polynomial regression model}
\label{sec: piecewise polynomial regression model}

This section provides an overview of the piecewise polynomial
regression model in a context of curves description and briefly
recalls the algorithm used to estimate its parameters.

The piecewise polynomial regression is a  segmentation method that
partitions the data into $K$ segments (or regimes). Generally used
to model a single curve or time series
\cite{McGee,brailovsky,ferrari1,chamroukhiESANN2009}, the piecewise
polynomial regression model can be used to model a set of curves
\cite{HugueneyEtAl:ESANN2009}. The parameters estimation is
performed using a dynamic programming procedure
\cite{bellman,fisher,yveslechevalier90} due to the additivity of the
optimized cost function over the $K$ segments.

Let $\mathcal{X}$ be a training set of $n$ curves
$\{\bx_1,\ldots,\bx_n\}$ where each curve $\bx_i$ consists of $m$
measurements $(x_{i1},\ldots,x_{im})$ observed at the time points
$(t_1,\ldots,t_m)$. In the following, the term ``curves size'' will be used to define $m$.
The piecewise regression model assumes that the curves $\mathcal{X}$ incorporate $K$ polynomial regimes defined on $K$ intervals whose bounds indexes can be denoted by $\bsgamma = (\gamma_1,\ldots,\gamma_{K+1})$ with $\gamma_1=0$ and $\gamma_{K+1}=m$. This defines a partition of $\mathcal{X}$ into $K$ segments of curves $\{\bsX_1,\ldots,\bsX_K\}$ of lengths $m_1,\ldots,m_K$ respectively where $\bsX_{k}=(\bx_{1}^k,\ldots,\bx_{n}^k)^T$ is the $ n m_k  \times 1$ vector of the elements in the $k$th segment for the $n$ curves with $\bx_{i}^k = \{x_{ij}|j\in I_k\}$ is the set of elements in segment $k$ of the $i$th curve whose indexes are $I_k = (\gamma_{k},\gamma_{k+1}]$. Therefore, for each curve $\bx_i$ $(i=1,\ldots,n)$, the piecewise polynomial regression model can be defined as follows:
\begin{equation}
\forall j=1,\ldots,m, \quad x_{ij} = \bsbeta^T_k\bsr_{j} +\sigma_k \epsilon_{ij}, \quad \epsilon_{ij} \sim \mathcal{N}(0,1) ,
\label{eq.piecewise regression model}
\end{equation}
where $k$ satisfies $j \in I_k$, $\bsbeta_k$ is the $p+1$ dimensional coefficients vector of a $p$ degree polynomial associated with the $k$th segment with $k \in \{1,\ldots,K\}$, $\bsr_{j}=(1,t_j,t_j^2\ldots,t_j^p)^T$ is the time dependent $p+1$ dimensional covariate vector associated with $\bsbeta_{k}$. As in classical regression models, the $\epsilon_{ij}$ are assumed to be independent random variables distributed according to a standard Gaussian distribution representing the additive noise. 

The model parameters can be denoted by $(\bspsi,\bsgamma)$ where \linebreak $\bspsi=(\bsbeta_1,\ldots,\bsbeta_K,\sigma_1^2,\ldots,\sigma_K^2)$ is the set of polynomial coefficients and noise variances, and $\bsgamma=(\gamma_{1},\ldots,\gamma_{K+1})$ is the set of the transition points.

\subsection{Maximum likelihood estimation for the piecewise polynomial regression model}
\label{ssec: maximum likelihood estimation for the piecewise polynomial regression model}

The estimation of the parameter vector $(\bspsi,\bsgamma)$ is performed by maximum likelihood.
As in classical model-based learning problems where each observation is described by a feature vector \cite{hastieTibshiraniFreidman_book_2009}, we assume that the curves sample $\{\bx_1,\ldots,\bx_n\}$ is independent. Within a segment $I_k$, the independence of the noises $\epsilon_{ij}$ ($j \in I_k$) involves the independence of $x_{ij}$ ($j \in I_k$) conditionally on $t_j$ ($j \in I_k$). Thus, according to the model (\ref{eq.piecewise regression model}), it can be proved that the observation $x_{ij}$, given the segment $k$, has
a Gaussian distribution with mean $\bsbeta_k^T\bsr_j$ and variance
$\sigma_k^2$. Therefore, the distribution of a curve $\bx_i$ is
given by:
\begin{equation}
p(\bx_i;\bspsi,\bsgamma) = \prod_{k=1}^K \prod_{j \in I_k}\mathcal{N}\left(x_{ij};\bsbeta_k^T\bsr_j,\sigma_k^2\right),
\label{eq. single curve distribution in the piecewise regression model}
\end{equation}
 and the log-likelihood of the parameter vector $(\bspsi,\bsgamma)$ characterizing the piecewise regression model, given the curves sample $\{\bx_1,\ldots,\bx_n\}$ is then written as follows:
\begin{eqnarray}
\!\!\!\! \!\!\!\! L(\bspsi,\bsgamma;\mathcal{X})
&\!\!\!\!=\!\!\!\! & \sum_{k=1}^K\sum_{i=1}^n \sum_{j\in I_k}\log  \mathcal{N}\left(x_{ij};\bsbeta_k^T\bsr_j,\sigma_k^2\right) \nonumber \\
&\!\!\!\!=\!\!\!\! &-\frac{1}{2}\sum_{k=1}^K \!\! \left[\frac{1}{\sigma_k^2}\sum_{i=1}^n \sum_{j\in I_k}\left(x_{ij}-\bsbeta_k^{T}\bsr_j \right)^2 \! + \! n m_k \log \sigma_k^2 \right]\! -  \! \frac{n m}{2} \log 2\pi,
\end{eqnarray}
where $m_k$ is the cardinal number of $I_k$.

Maximizing this log-likelihood is equivalent to minimizing, with respect to $\bspsi$ and $\bsgamma$, the criterion
\begin{eqnarray}
J(\bspsi,\bsgamma) &=& \sum_{k=1}^K \left[\frac{1}{\sigma_k^2}\sum_{i=1}^n \sum_{j\in I_k}\left(x_{ij}-\bsbeta_k^{T}\bsr_j \right)^2 + n m_k \log \sigma_k^2 \right].
\label{eq.piecewise_reg criterion J}
\end{eqnarray}

The next section shows how the parameters $\bspsi$ and $\bsgamma$ can be estimated using dynamic programming.

\subsection{Parameter estimation for the piecewise regression model by the Fisher algorithm }
\label{ssec: fisher's algorithm}

Fisher algorithm is based on a dynamic programming procedure that provides the optimal segmentation of the data by minimizing an additive criterion \cite{fisher,yveslechevalier90,brailovsky}. It can be used to minimize (\ref{eq.piecewise_reg criterion J}) with respect to $\bspsi$ and $\bsgamma$ or equivalently to minimize (\ref{dynamic programming criterion C}) with respect to $\bsgamma$:
\begin{eqnarray}
C(\bsgamma) &=& \min \limits_{\substack {\bspsi}} J(\bspsi,\bsgamma) \nonumber \\
&=& \sum_{k=1}^K \min \limits_{\substack {(\bsbeta_k,\sigma_k^2)}} \left[\frac{1}{\sigma_k^2}\sum_{i=1}^n\sum_{j=\gamma_k+1}^{\gamma_{k+1}}\left(x_{ij}-\bsbeta_k^{T}\bsr_j \right)^2 + n m_k \log \sigma_k^2 \right] \nonumber \\
& = & \sum_{k=1}^K\left[ \frac{1}{\hat{\sigma}_k^2} \sum_{i=1}^n \sum_{j=\gamma_k+1}^{\gamma_{k+1}}(x_{ij}-\hat{\bsbeta}_k^{T}\bsr_j)^2 + n m_k \log \hat{\sigma}_k^2 \right] ,
\label{dynamic programming criterion C}
\end{eqnarray}
with
\begin{equation}
{\hat{\bsbeta}}_k = \arg \min \limits_{\substack{\bsbeta_k}} \sum_{i=1}^n \sum_{j=\gamma_k+1}^{\gamma_{k+1}}(x_{ij}-\bsbeta_k^{T}\bsr_j)^2 = (\bM_k^T\bM_k)^{-1}\bM_k^T\bsX_k,
\label{eq. estimation of betak in piecewise reg}
\end{equation}
and
\begin{equation}
\hat{\sigma}_k^2 =\frac{1}{ n m _k} \sum_{i=1}^n\sum_{j=\gamma_k+1}^{\gamma_{k+1}} (x_{ij}-{\hat{\bsbeta}}_k^T \bsr_j)^2,
\label{eq. estimation of sigmak in piecewise reg}
\end{equation}
where 
$\bM_k=\left[\begin{array}{c}
\Phi_k\\ \vdots\\ \Phi_k \end{array}\right]$ is the $ n m_k \times (p+1) $ regression matrix of the segment $k$  for all the curves  and
$$\Phi_k=\left[\begin{array}{ccccc}
1&t_{\gamma_{k}+1}& t_{\gamma_{k}+1}^2&\ldots&t_{\gamma_{k}+1}^p\\
1&t_{\gamma_{k}+2}& t_{\gamma_{k}+2}^2&\ldots&t_{\gamma_{k}+2}^p\\
\vdots&\vdots&\vdots&\vdots&\vdots\\
1&t_{\gamma_{k+1}}& t_{\gamma_{k+1}}^2&\ldots&t_{\gamma_{k} +1}^p
\end{array}\right]$$
is the $ m_k\times(p+1) $ regression matrix for the segment $k$ for each curve.

We can see that the criterion $C(\bsgamma)$ is the sum of the cost \linebreak $\frac{1}{\hat{\sigma}_k^2} \sum_{i=1}^n \sum_{j=\gamma_k+1}^{\gamma_{k+1}}(x_{ij}-\hat{\bsbeta}_k^{T}\bsr_j)^2 + n m_k \log \hat{\sigma}_k^2 $ over the $K$ segments. The additivity of this criterion means it can be optimized globally using a dynamic programming procedure \cite{bellman,yveslechevalier90}. Dynamic programming considers that an optimal partition of the data into $K$ segments is the union of an optimal partition into $K-1$ segments and one segment. Thus, by denoting by $C_1(a,b)$ the optimal cost within one segment whose elements indexes are $(a,b]$ with $0\leq a <b \leq m$, the optimal costs $C_k(a,b)$ for a partition into $k$ segments, $k=2,\ldots,K$, is recursively computed as follows:
\begin{eqnarray}
\left \{ \begin{tabular}{lll}
$C_1(a,b)$&=&$\min \limits_{\substack {(\bsbeta,\sigma^2)}} \left[\frac{1}{\sigma^2}\sum_{i=1}^n\sum_{j=a+1}^{b}\left(x_{ij}-\bsbeta^{T}\bsr_j \right)^2 + n (b-a) \log \sigma^2 \right]$\\
&=&$\frac{1}{\hat{\sigma}^2}\sum_{i=1}^{n}\sum_{j=a+1}^{b} (x_{ij}-\hat{\bsbeta}^{T}\bsr_j)^2 + n (b-a) \log \hat{\sigma}^2$\\
&&\\
$C_k \left(a,b\right)$& = &$\min \limits_{\substack {a\leq h\leq b}} \left[C_{k-1}\left(a,h\right) + C_1\left(h+1,b\right)\right]$\quad for $k=2,\ldots,K.$
\end{tabular}\right.
\label{cost_matrix_and_recursive_formula}
\end{eqnarray}
where $\hat{\bsbeta}$ and $\hat{\sigma}^2$ are computed respectively according to the equations (\ref{eq. estimation of betak in piecewise reg}) and (\ref{eq. estimation of sigmak in piecewise reg}) by replacing $(\gamma_k,\gamma_{k+1}]$ by $(a,b]$, $m_k$ by $(b-a)$ and $\hat{\bsbeta}_k$ by $\hat{\bsbeta}$. Thus, the algorithm works as follows:
\subsubsection*{\textbf{Step 1.} (Initialization)} This step consists of computing the cost matrix $C_1(a,b)$ for one segment $(a,b]$ for $0\leq a <b \leq m$ using (\ref{cost_matrix_and_recursive_formula}).
\subsubsection*{\textbf{Step 2.} (Dynamic programming procedure)}
This step consists of recursively computing the optimal cost $C_k(a,b)$ for $k=2,\ldots,K$ and $0\leq a <b \leq m$ using (\ref{cost_matrix_and_recursive_formula}).
\subsubsection*{\textbf{Step 3.} (Finding the optimal partition)}
The optimal partition can be deduced from the optimal costs $C_k(a,b)$. (For more details see appendix A of \cite{brailovsky}).

This algorithm has a time complexity of $\mathcal{O}(Kp^2 n^2m^2)$ which
can be computationally expensive for large sample sizes.

\subsection{Curves approximation and classification with the piecewise regression model}

\subsubsection{Curves approximation}

Once the model parameters are estimated, the curves approximation derived from the piecewise polynomial regression model is given by \linebreak $\hat{x}_{ij}= \sum_{k=1}^K \hat{z}_{jk} \hat{\bsbeta}^T_k \bsr_j$, $\forall$ $i=1,\ldots,n$ where
$\hat{z}_{jk}=1$ if $j \in (\hat{\gamma}_k,\hat{\gamma}_{k+1}]$ and $\hat{z}_{jk}=0$ otherwise. The vectorial formulation of the curves approximation
$\hat{\mathcal{X}}$ can be written as:
\begin{equation}
\hat{\mathcal{X}} = \sum^{K}_{k=1} \hat{Z}_{k} \bT \hat{\bsbeta}_{k},
\end{equation}
where $\hat{Z}_{k}$ is a diagonal matrix whose diagonal elements are
$(\hat{z}_{1k} ,\ldots,\hat{z}_{mk} )$, and
$$\bT=\left[\begin{array}{ccccc}
1&t_{1}&t_{1}^2&\ldots&t_{1}^p\\
1&t_{2}&t_{2}^2&\ldots&t_{2}^p \\
\vdots&\vdots&\vdots&\vdots&\vdots\\
1&t_{m}&t_{m}^2&\ldots&t_{m}^p\end{array}\right]$$ is the $m \times
(p+1)$ regression matrix.

\subsubsection{Curve classification}
\label{ssec: curve classification with the piecewise approach}

This section presents the discrimination rule which can be derived from the piecewise polynomial regression approach to classify curves into predefined classes.

Let us denote by $C_i$ the class label of the curve $\bx_i$,  which
takes its values in the finite set $\{1,\ldots,G\}$  where $G$ is
the number of classes. Given a labeled training set of curves, the
parameter vectors
$(\hat{\bspsi}_1,\hat{\bsgamma}_1)$,...,$(\hat{\bspsi}_G,\hat{\bsgamma}_G)$
for the $G$ classes are estimated by the dynamic programming
procedure. Once the classes parameters are estimated, a new acquired
curve $\bx_i$ is assigned to the class $\hat{g}$ that maximizes the
posterior probability that $\bx_i$ belongs to the class $g$, ($g=1,\ldots,G$):
\begin{equation}
\hat{g}=\arg \max \limits_{\substack{1\leq g\leq G}} p(C_i=g|\bx_i;\hat{\bspsi}_g,\hat{\bsgamma}_g),
\label{MAP rule}
\end{equation}
where
\begin{equation}
p(C_i=g|\bx_i;\hat{\bspsi}_g,\hat{\bsgamma}_g)=\frac{p(C_i=g)p(\bx_i|C_i=g;\hat{\bspsi}_g,\hat{\bsgamma}_g)}{\sum_{g'=1}^{G}p(C_i=g')p(\bx_i|C_i=g';\hat{\bspsi}_{g'},\hat{\bsgamma}_{g'})} ,
\label{eq. map rule picewise regression}
\end{equation}
$p(C_i=g)$ being the proportion of the class $g$ in the training
database and $p(\bx_i|C_i=g;\hat{\bspsi}_g,\hat{\bsgamma}_g)$ the
conditional density of $\bx_i$ given the class $g$ defined by
equation (\ref{eq. single curve distribution in the piecewise
regression model}). The parameters
$(\hat{\bspsi}_g,\hat{\bsgamma}_g)$ represent the  maximum
likelihood estimates of $(\bspsi_g, \bsgamma_g)$.

\section{The proposed regression model with a hidden logistic process}
\label{sec: proposed regression model}

Although the piecewise regression model described in the previous
section is based on the global optimization of a likelihood
criterion, it is naturally tailored for curves presenting abrupt
changes since the obtained curve segmentation is hard. Moreover, it
is well known that the dynamical programming procedure is
computationally expensive for large sample sizes. This section
presents the proposed regression model based on a hidden logistic
process for functional data modeling. The flexibility of this model
allows for modeling of curves with abrupt or smooth changes in
regime.

\subsection{The global regression model}

In the proposed model, each curve $\bx_i$ from the set $\{\bx_1,...,\bx_n\}$ is assumed to be generated by the following regression model with a discrete hidden process $\bz=(z_1,\ldots,z_m)$:
\begin{equation}
\forall j=1,\ldots,m, \quad x_{ij}= \bsbeta^T_{z_{j}}\bsr_{j} + \sigma_{z_{j}}\epsilon_{ij}, \quad\epsilon_{ij} \sim \mathcal{N}(0,1),
\label{eq.regression model hidden logistic process}
\end{equation}
where $z_{j} \in\{1,\ldots,K \}$ is a hidden discrete variable representing the label of the polynomial regression model generating $x_{ij}$. This model can be reformulated in a matrix form by
\begin{equation}
\bx_i = \sum^{K}_{k=1} Z_k (\bT \bsbeta_{k} + \sigma_k \boldsymbol{\epsilon}_i), \quad \bsepsilon_{i} \sim \mathcal{N}(\boldsymbol{0},\boldsymbol{I}_m),
\label{ecriture_vectorielle_du_modele}
\end{equation}
where $Z_k$ is the $ m \times m $ diagonal matrix whose diagonal elements are \linebreak $(z_{1k},\ldots,z_{mk})$, with $z_{jk}=1$ if $z_{j}=k$ (i.e if $x_{ij}$ is generated by the $k$th regression model) and $z_{jk}=0$ otherwise. The variable $\boldsymbol{\epsilon}_i=(\epsilon_{i1},\ldots,\epsilon_{im})^T$ is a $m\times 1$ noise
vector distributed according to a Gaussian density with zero mean
and identity covariance matrix.

The next section defines the probability distribution of the process \linebreak $\bz=(z_1,\ldots,z_m)$ that allows the switching from one regression model to another.

\subsection{The hidden logistic process}
\label{ssec: process}

The proposed hidden logistic process assumes that the variables $z_{j}$, given the vector $\bst=(t_1,\ldots,t_m)$, are generated independently according to the multinomial distribution {\small$\mathcal{M}(1,\pi_{j1}(\bw),\ldots,\pi_{jK}(\bw))$}, where
\begin{equation}
\pi_{jk}(\bw)= p(z_{j}=k;\bw)=\frac{\exp{(\bsw_{k0} +\bsw_{k1}t_j)}}{\sum_{\ell=1}^K\exp{(\bsw_{\ell 0} + \bsw_{\ell 1} t_j)}},
\label{eq. multinomial logistic function}
\end{equation}
is the logistic transformation of a linear function of the time point $t_j$,  \linebreak $\bsw_{k}=(\bsw_{k0},\bsw_{k1})^T$ is the $2$ dimensional coefficients vector for the $k$th component of (\ref{eq. multinomial logistic function}) %associated to the covariate $(1,t_j)^T$
and $\bw = (\bsw_1,\ldots,\bsw_K)$. Thus, given the vector $\bst=(t_1,\ldots,t_m)$, the distribution of $\bz$ can be written as:
\begin{equation}
p(\bz;\bw)=\prod_{j=1}^m \prod_{k=1}^K \left(\frac{\exp{(\bsw_{k0} +\bsw_{k1}t_j)}}{\sum_{\ell=1}^K\exp{(\bsw_{\ell 0} + \bsw_{\ell 1} t_j)}}\right)^{z_{jk}}.
\label{eq.hidden logistic process}
\end{equation}

The relevance of the logistic transformation in terms of flexibility
of transition can be illustrated through simple examples with $K=2$
components. In this case, only the probability $\pi_{j1}(\bw)=
\frac{exp(\bsw_{10} + \bsw_{11}t_j)}{1+exp(\bsw_{10} +
\bsw_{11}t_j)}$ should be described, since
$\pi_{j2}(\bw)=1-\pi_{j1}(\bw)$. The variation of the proportions
$\pi_{jk}(\bw)$ over time, in relation to the parameter $\bsw_k$, is
illustrated by an example of 2 classes where we use the
parametrization $\bsw_k =\lambda_k (\alpha_k, 1)^T$, with
$\lambda_k= \bsw_{k1}$ and $\alpha_k = \frac{\bsw_{k0}}{\bsw_{k1}}
\cdot$

\begin{figure*}[!h]
\centering
\begin{tabular}{cc}
\includegraphics[height=5.5cm,width=6.5cm]{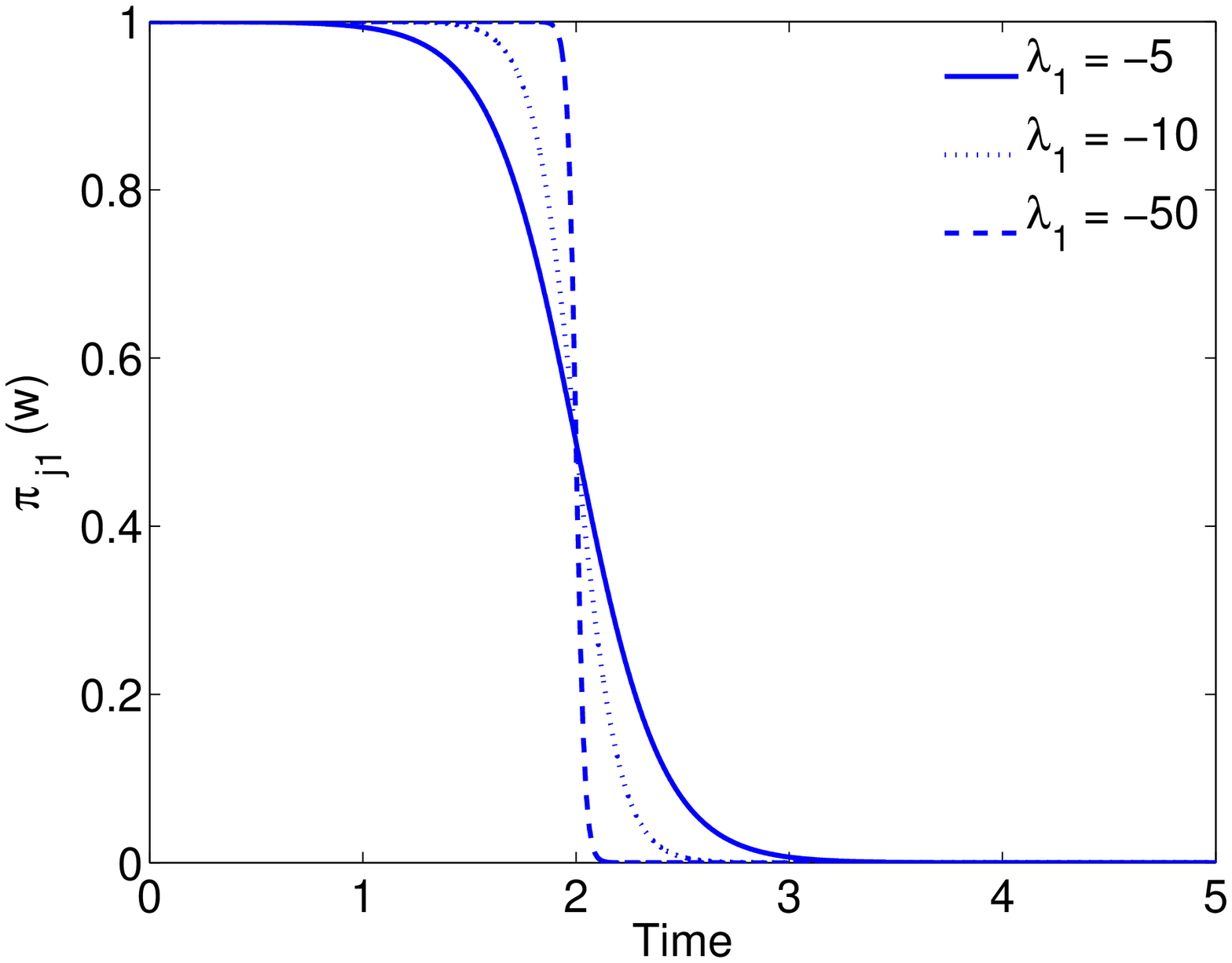} &
\includegraphics[height=5.5cm,width=6.5cm]{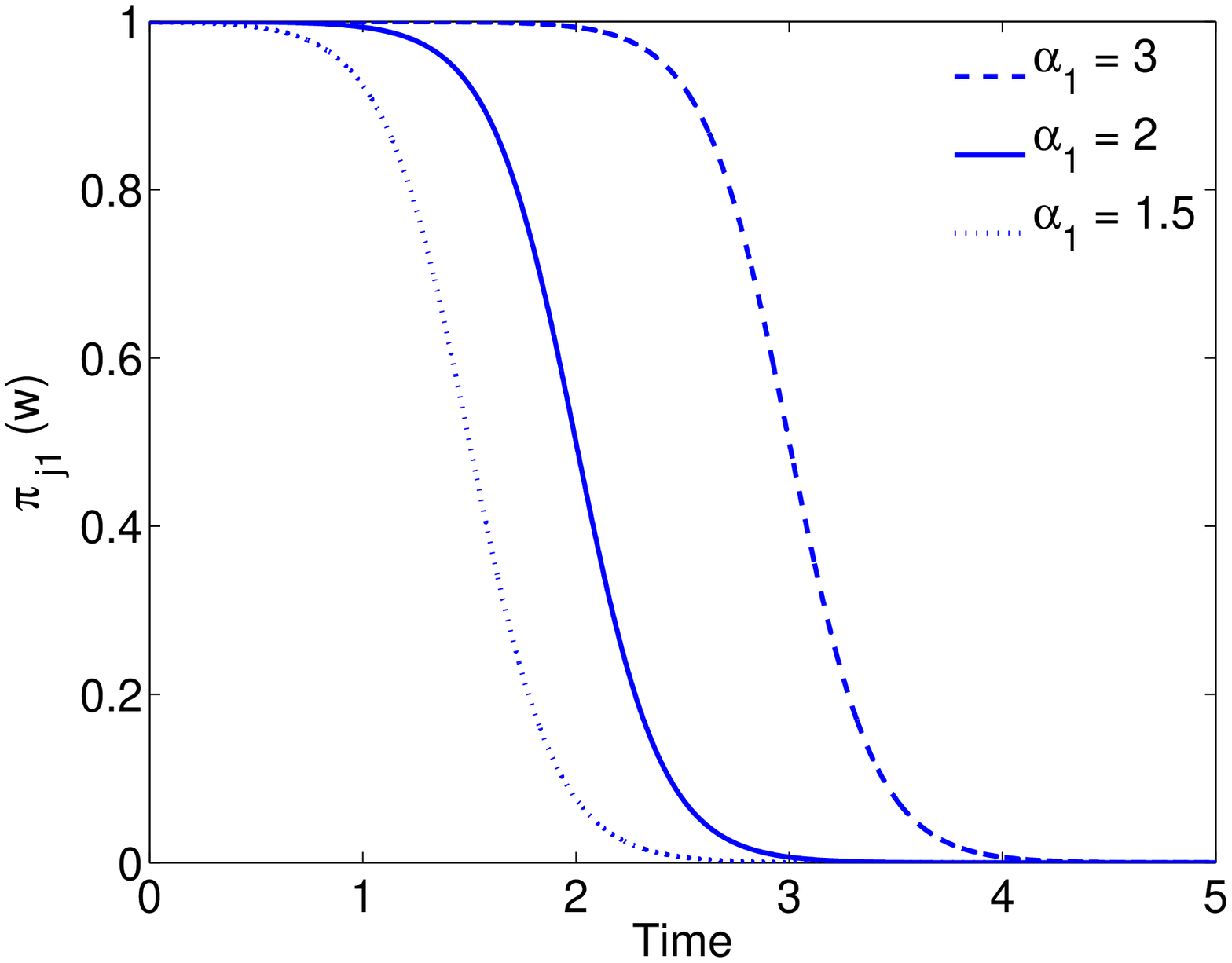}\\
\small{(a)} & \small{(b)}
\end{tabular}
\caption{Variation of $\pi_{j1}(\bw)$ over time for a dimension $q=1$ of $\bsw_1$ and (a) different values of $\lambda_1 = \bsw_{11}$ with $\alpha_1 = -2$ and (b) different values of $\alpha_1 = \frac{\bsw_{10}}{\bsw_{11}}$ with $\lambda_1=-5$.}
\label{logistic_function_k=2_q=1}
\end{figure*}

As shown in Fig. \ref{logistic_function_k=2_q=1} (a), the parameter $\lambda_k$ controls the quality of transitions between the regression models, the higher absolute value of $\lambda_k$, the more abrupt the transition between the $z_j$, while the parameter $\alpha_k$ controls the transition time point via the inflexion point of the curve (see Fig. \ref{logistic_function_k=2_q=1} (b)).

In this particular regression model, the variable $z_{j}$ controls the switching from one regression model to another of $K$ regression models within the curves at each time $t_j$. The use of the logistic process for modeling the sequence of variables $z_j$ allows for modeling both abrupt and smooth transitions between the regimes within the curves, unlike the piecewise regression model which is adapted only for regimes with abrupt transitions.
\subsection{The generative model of curves}
\label{ssec: the generative model of curves}
The generative model of $n$ curves from a fixed parameters $(\bw,\bsbeta_k,\sigma^2_k)$ for $k=1,\ldots,K$ consists of 2 steps:
\begin{itemize}
\item generate the hidden process $\bz=(z_1,\ldots,z_m)$ according to the multinomial distribution $z_{j} \sim {\small \mathcal{M}(1,\pi_{j1}(\bw),\ldots,\pi_{jK}(\bw))}$,
\item for $i=1,\ldots,n$ and for $j=1,\ldots,m$: generate each observation $x_{ij}$ according to the Gaussian distribution $\mathcal{N}(\cdot;\bsbeta^{T}_{z_{j}}\bsr_j,\sigma^{2}_{z_{j}})$.
\end{itemize}

\subsection{Parameter estimation}
\label{ssec: parameter estimation HLP regression}

From the model (\ref{eq.regression model hidden logistic process}), it can be proved that, conditionally on a regression model $k$, $x_{ij}$ is distributed according to a normal density with mean $\bsbeta^T_k\bsr_{j}$ and variance $\sigma^2_k$. Thus, it can be proved that $x_{ij}$ is distributed according to the normal mixture density
\begin{equation}
p(x_{ij};\bstheta)=\sum_{k=1}^K\pi_{jk}(\bw)\mathcal{N}\big(x_{ij};\bsbeta^T_k\bsr_{j},\sigma^2_k\big) ,
\label{melange}
\end{equation}
where $\bstheta=(\bw,\bsbeta_1,\ldots,\bsbeta_K,\sigma^2_1,\ldots,\sigma^2_K)$
is the parameter vector to be estimated. The parameter $\bstheta$ is estimated by the maximum likelihood method.

As in the piecewise polynomial regression model, we assume that the curves sample
$\mathcal{X}=\{\bx_1,\ldots, \bx_n\}$ is independent. The independence of the $\epsilon_{ij}$'s ($j=1,\ldots,m$) involves the independence of the $x_{ij}$'s ($j=1,\ldots,m$) conditionally on the time vector $\bst =(t_1,\ldots,t_m)$. It should be noticed that the temporal dependence between the underlying segments is controlled by the logistic distribution. The distribution of $\bx_i$ is then written as:
\begin{equation}
p(\bx_i;\bstheta) = \prod_{j=1}^m \sum_{k=1}^K\pi_{jk}(\bw)\mathcal{N}\big(x_{ij};\bsbeta^T_k\bsr_{j},\sigma^2_k\big).
\label{eq. single curve distribution in the proposed model}
\end{equation}
Therefore, the log-likelihood of $\bstheta$ is written as:
\begin{eqnarray}
L(\bstheta;\mathcal{X})&=&\log p(\bx_1,\ldots,\bx_n;\bstheta)\nonumber\\
&=&\sum_{i=1}^{n}\sum_{j=1}^{m}\log\sum_{k=1}^K \pi_{jk}(\bw)\mathcal{N}\big(x_{ij};\bsbeta^T_k\bsr_j,\sigma^2_k\big) .
\end{eqnarray}
The direct maximization of this likelihood is not straightforward, we use a dedicated Expectation Maximization
(EM) algorithm \cite{dlr,mclachlanEM} to perform the maximization.

\subsection{The dedicated EM algorithm}
\label{ssec. EM algorithm}

The proposed EM algorithm starts from an initial parameter $\bstheta^{(0)}$ and alternates the two following steps until convergence:
\subsubsection*{\textbf{E Step (Expectation)}}
This step consists in computing the expectation of the complete log-likelihood $CL(\bstheta;\mathcal{X},\bz) = \log p(\mathcal{X},\bz;\bstheta)$ given the observed data $ \mathcal{X}$ and the current value $\bstheta^{(q)}$ of the parameter $\bstheta$ ($q$ being the current iteration):
\begin{equation}
Q(\bstheta,\bstheta^{(q)})=E\left[CL(\bstheta;\mathcal{X},\bz)|\mathcal{X};\bstheta^{(q)}\right].
\end{equation}
This step simply requires the computation of the posterior probabilities
\begin{eqnarray}
\!\! \tau^{(q)}_{ijk} &=& p(z_{jk}=1|x_{ij};\bstheta^{(q)}) = \frac{\pi_{jk}(\bw^{(q)})\mathcal{N}(x_{ij};\bsbeta^{T(q)}_k\bsr_{j},\sigma^{2(q)}_k)}
{\sum_{\ell=1}^K\pi_{j \ell}(\bw^{(q)})\mathcal{N}(x_{ij};\bsbeta^{T(q)}_{\ell}\bsr_{j},\sigma^{2(q)}_{\ell})}
\label{eq.tik}
\end{eqnarray}
that $x_{ij}$ originates from the $k$th regression model (see appendix for details).

\subsubsection*{\textbf{M step (Maximization)} }
In this step, the value of the parameter $\bstheta$ is updated by computing the parameter $\bstheta^{(q+1)}$ maximizing the conditional expectation $Q$ with respect to $\bstheta$.

Maximizing  $Q$ with respect to $\bsbeta_k$ ($k=1,\ldots,K$) consists in analytically solving a weighted least-squares problem. The estimates are given by:
\begin{equation}
{\bsbeta}_k^{(q+1)} = (\Lambda^T W_k^{(q)}\Lambda)^{-1}\Lambda^T W_k^{(q)} \mathcal{X},
\label{eq. estimation betaEM}
\end{equation}
where $ W_k^{(q)}$ is the $nm \times nm$ diagonal matrix whose diagonal elements are the posterior probabilities $(\tau_{11k}^{(q)},\ldots,\tau_{1mk}^{(q)},\ldots,\tau_{n1k}^{(q)},\ldots,\tau_{nmk}^{(q)})$ for the $k$th regression component and $\Lambda$ is the $nm\times (p+1)$ regression matrix for all the curves $\mathcal{X}$ such that:
$$\Lambda =\left[\begin{array}{c}
                    \bT \\
                    \vdots\\
                    \bT \\
                  \end{array}\right]
                  \mbox{ and } \mathcal{X}=\left[
                                \begin{array}{c}
                                  \bx_{1} \\
                                  \vdots \\
                                  \bx_{n} \\
                                \end{array} \right].$$

Maximizing  $Q$ with respect to $\sigma_k^2$ ($k=1,\ldots,K$) provides the following updating formula:
\begin{equation}
{\sigma}_k^{2(q+1)}= \frac{1}{n m_k^{(q)}}\sum_{i=1}^{n}\sum_{j=1}^{m} \tau^{(q)}_{ijk} (x_{ij}-{\bsbeta}_k^{T(q+1)}\bsr_j)^2,
\label{eq. estimation sigmaEM}
\end{equation}
where $m_k^{(q)}=\sum_{j=1}^{m} \tau^{(q)}_{ijk}$ is the cardinal number of the component $k$ estimated at iteration $q$ for each curve $\bx_i$ (see appendix for more details).

The maximization of $Q$ with respect to $\bw$ is a multinomial logistic regression problem weighted by $\tau^{(q)}_{ijk}$ which can be solved using a multi-class Iterative Reweighted Least Squares (IRLS) algorithm \cite{irls,chen99,krishnapuram,chamroukhi et al. NN2009}.

The proposed algorithm is performed with a time complexity of \linebreak $\mathcal{O}(NMnm K^3p^2)$, where $N$ is the number of iterations of the EM algorithm and $M$ is the average number of iterations required by the IRLS algorithm used in the maximization step at each iteration of the EM algorithm. Thus, the ratio between the time complexity of the piecewise polynomial regression model and the time complexity of the proposed regression model is $nm/NMK^2$. In practice, as illustrated in the computing time graphics (see Fig. \ref{fig. cputime_varying_m_and_n}), $nm$ is larger than $NM K^2$ since the number of regimes $K$ does not exceed 5 and a particular strategy is used to initialize the IRLS algorithm. This initialization consists in choosing an arbitrary value of the parameter $\bw$ only for the first iteration of the EM algorithm. For the other EM iterations, the IRLS loop starts with the parameter $\bw^{(q)}$ estimated at the $q$th iteration of the EM algorithm and provides $\bw^{(q+1)}$. This setting reduces the running time of the IRLS algorithm and thus reduces the running time of the EM algorithm.

\subsection{Model selection}

The optimal values of the pairs $(K,p)$ can be computed by using the Bayesian Information
Criterion (BIC) \cite{BIC} which is a penalized
likelihood criterion, defined by
\begin{equation}
\mbox{BIC}(K,p) = L(\hat{\bstheta};\mathcal{X}) - \frac{\nu(K,p)\log(nm)}{2},
\end{equation}
where $\nu(K,p) = K(p+4)-2$ is the number of parameters of the model and $L(\hat{\bstheta};\mathcal{X})$ is the log-likelihood obtained at convergence of the EM algorithm.  

\subsection{Curves approximation and classification with the proposed model}
\label{ssec: curves approximation and classification with the proposed model}

\subsubsection{Curves approximation}

With the proposed description approach, the set of curves belonging to the same class is approximated by a single curve. Each point of this curve is given by the expectation  $E(x_{ij};\hat{\bstheta})$  $\forall$ $i=1,\ldots,n$ and $\forall$ $j=1,\ldots,m$ where
\begin{eqnarray}
E(x_{ij};\hat{\bstheta}) &=& \int_{\IR}x_{ij} p(x_{ij};\hat{\bstheta})dx_{ij}\nonumber\\
 &=& \sum_{k=1}^K \pi_{jk}(\hat{\bw})\int_{\IR}x_{ij} \mathcal{N}\big(x_{ij};\hat{\bsbeta}^T_k\bsr_{j},\hat{\sigma}^2_k\big) dx_{ij}\nonumber\\
 &=& \sum_{k=1}^{K} \pi_{jk}(\hat{\bw})\hat{\bsbeta}^T_k \bsr_{j}, 
\end{eqnarray}
$\hat{\bstheta} = (\hat{\bw},\hat{\bsbeta}_1,\ldots,\hat{\bsbeta}_K,\hat{\sigma}^2_1,\ldots,\hat{\sigma}^2_K)$ being the parameter vector obtained at convergence of the algorithm.
The matrix formulation of the curves approximation $\hat{\mathcal{X}}$ is given by:
\begin{equation}
\hat{\mathcal{X}} = \sum_{k=1}^K \hat{\mathcal{W}}_{k} \bT \hat{\bsbeta}_{k},
\label{eq. curve expectation with the proposed approach}
\end{equation}
where $\hat{\mathcal{W}}$ is a diagonal matrix whose diagonal elements are the proportions $( \pi_{1k}({\hat{\bw}}),\ldots,\pi_{mk}({\hat{\bw}})) $ associated with the $k$th regression model.

\subsubsection{Curve classification}
\label{ssec: curve classification with the proposed approach}

Basing on the proposed curves modeling approach, a curve discrimination rule can be derived. Given a labelled training set of curves, the parameters $\bstheta_1 ,\ldots,  \bstheta_G$ of the $G$ classes are first estimated by applying the proposed description approach to each class of curves. This approach is generally used for supervised learning of generative models. Whereas, for the discriminative approaches, which directly estimate the decision boundaries, the parameters of each class are not estimated independently from the other classes.

Once the classes parameters are estimated by the EM algorithm, a new acquired curve $\bx_i$ is assigned to the class $\hat{g}$, as described in section \ref{ssec: curve classification with the piecewise approach}, by the MAP rule:
\begin{equation}
\hat{g}=\arg \max \limits_{\substack{1\leq g\leq G}} p(C_i=g|\bx_i;\hat{\bstheta}_g),
\label{MAP rule}
\end{equation}
where
\begin{equation}
p(C_i=g|\bx_i;\hat{\bstheta}_g)=\frac{p(C_i=g)p(\bx_i|C_i=g;\hat{\bstheta}_g)}{\sum_{g'=1}^{G}p(C_i={g'})p(\bx_i|C_i={g'};\hat{\bstheta}_{g'})} ,
\label{eq. map rule proposed model}
\end{equation}
$p(\bx_i|C_i=g;\hat{\bstheta}_g)$ being the conditional density of $\bx_i$ given the class $g$ defined by equation (\ref{eq. single curve distribution in the proposed model}). The parameter vector $\hat{\bstheta}_g =(\hat{\bw}_g,\hat{\bsbeta}_{1g},\ldots,\hat{\bsbeta}_{Kg},\hat{\sigma}^2_{1g},\ldots,\hat{\sigma}^2_{Kg})$ is the maximum likelihood estimate of ${\bstheta}$ for the class $g$.

\section{Experimental study}
\label{sec: experiments}

This section is devoted to an evaluation of the proposed approach in terms of curves description and classification, using simulated data sets and real data sets. For this purpose, the proposed approach was compared with the piecewise polynomial regression approach. Two evaluation criteria  were used. 
\begin{itemize}
\item The first criterion is the mean square error between the true simulated curve without noise and the estimated curve given by:
\begin{itemize}
\item $\hat{x}_{ij}=\sum_{k=1}^{K} \pi_{jk}(\hat{\bw})\hat{\bsbeta}^T_k \bsr_{j}$ for the proposed model;
\item $\hat{x}_{ij}=\sum_{k=1}^{K}\hat{z}_{jk}\hat{\bsbeta}^T_{k}\bsr_{j}$ for the piecewise polynomial regression model.\end{itemize}
The mean square error criterion is computed by the formula \linebreak $\frac{1}{n m}\sum_{i=1}^{n} \sum_{j=1}^{m}[E(x_{ij};\bstheta)-\hat{x}_{ij}]^2$,
$\bstheta$ being the true parameter vector. It is used to assess the models with regard to curves modeling.

\item  The second criterion is the curves misclassification error rate computed by a $5$-fold cross-validation procedure.

\end{itemize}

\subsection{Evaluation in terms of curves modeling}
\label{ssec: evaluation in terms of curves modeling}

Three experiments were performed to evaluate the proposed approach in terms of curves modeling:
\begin{itemize}
\item  the first experiment aims at observing the effect of the smoothness level of transitions on estimation quality. The smoothness level of transitions was tuned by means of the term $\lambda_k=\bsw_{k1}$ seen in section \ref{ssec: process} and Fig. \ref{logistic_function_k=2_q=1} (a). Each simulated sample of curves consisted of $n=10$ curves with a curves size $m=100$. The simulated curves consisted of three constant polynomial ($K=3, p=0$)  with transition time points at $1$ and $3$ seconds. Each simulated curve consisted in a mean curve corrupted by an additive uniform zero-mean Gaussian noise with a standard deviation $\sigma=2$. The $j$th point of the mean curve is  given by $\sum_{k=1}^K \pi_{jk}(\bw) \bsbeta^T_{k} \bsr_j$. The set of simulation parameters $\{\bsbeta_k,\bsw_k\}$ for this experiment is given in Table \ref{table. parameters of simulation of experiment 1}. We have considered decreasing values of $|\lambda_k|$, which correspond to increasing values of the smoothness level of transitions (see Table \ref{table. levels of smoothness}). Fig. \ref{fig. curves with varying smoothness level} (a) shows the true denoised curves for the $10$ smoothness levels of transitions and Fig. \ref{fig. curves with varying smoothness level} (b) shows an example of simulated curves for a fixed smoothness level.

\begin{table}[!h]
\centering
\small
\begin{tabular}{|ll|}
\hline
$\bsbeta_1=0$ &$\bsw_1=[3341.33,-1706.96]$ \\
$\bsbeta_2=10$ & $\bsw_2=[2436.97,-810.07]$ \\
$\bsbeta_3=5$ & $\bsw_3=[0,0]$ \\
 \hline
\end{tabular}
\caption{Simulation parameters for experiment 1.}
\label{table. parameters of simulation of experiment 1}
\end{table}
\begin{table*}[!h]
\scriptsize
\centering
\begin{tabular}{|lllllllllll|}
\hline
Smoothness & & & & & & & & & & \\
level of &1 &2 &3 &4 &5 &6 &7 &8 &9 &10\\
transitions & & & & & & & & & & \\
\hline
$|\lambda_k|$& $\frac{|\bsw_{k1}|}{1}$ & $\frac{|\bsw_{k1}|}{2}$ & $\frac{|\bsw_{k1}|}{5}$ & $\frac{|\bsw_{k1}|}{10}$ & $\frac{|\bsw_{k1}|}{20}$ & $\frac{|\bsw_{k1}|}{40}$ & $\frac{|\bsw_{k1}|}{50}$ & $\frac{|\bsw_{k1}|}{80}$& $\frac{|\bsw_{k1}|}{100}$ & $\frac{|\bsw_{k1}|}{125}$\\
\hline
\end{tabular}
\caption{The different smoothness levels from abrupt transitions to smooth transitions for the situations shown in Fig. \ref{fig. curves with varying smoothness level} (a).}
\label{table. levels of smoothness}
\end{table*}
\begin{figure*}[!h]
\centering
\begin{tabular}{cc}
\includegraphics[width=6.6cm]{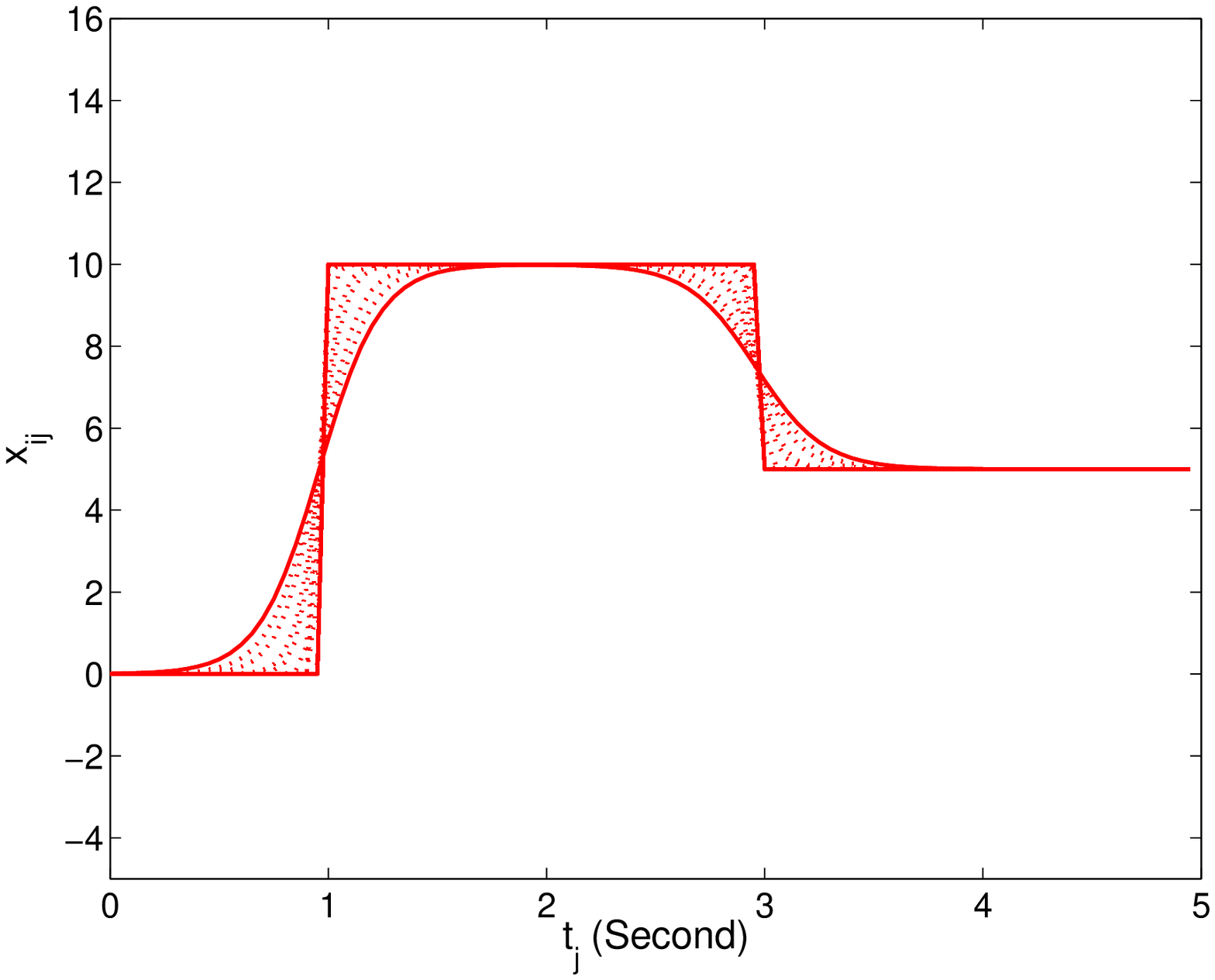}&
\includegraphics[width=6.6cm]{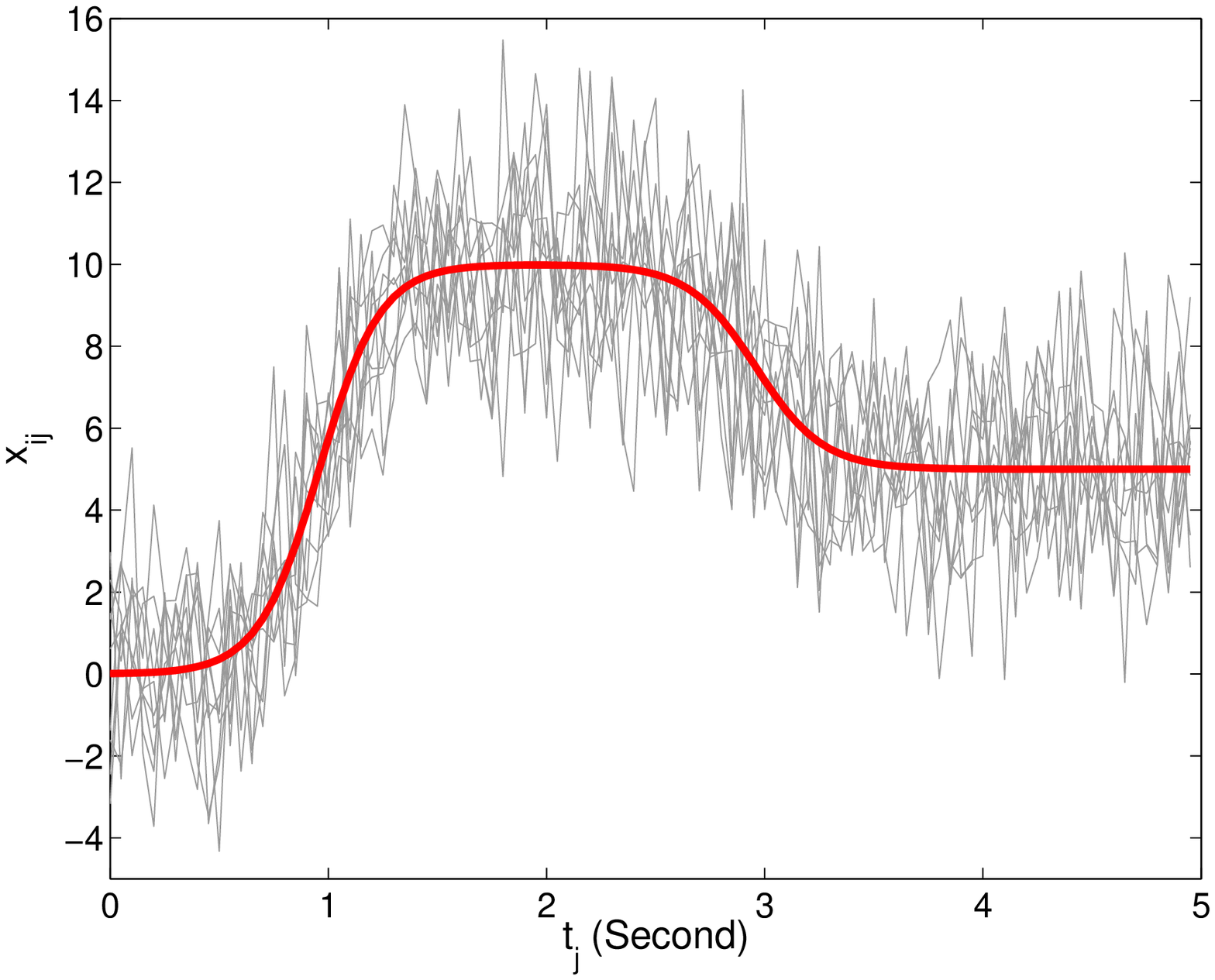}\\
\small{(a)}&\small{(b)}\\
\end{tabular}
\caption{The true denoised curves from abrupt transitions to smooth transitions for the first experiment (a) and an example of simulated curves ($n=10$, $m=100$) with a smoothness level of transition corresponding to the level 8 in Table \ref{table. levels of smoothness}.}
\label{fig. curves with varying smoothness level}
\end{figure*}

\item the second experiment aims at observing the effect of the sample size $n$  on estimation quality. It varies from $10$ to $100$ by step of $10$, and the  curves size was set to $m=100$.

\item the third experiment aims at observing the effect of the curves size $m$ on estimation quality. It varied from $100$ to $1000$ by step of $100$ for a fixed number of curves $n=50$.
\end{itemize}

\begin{figure}[!h]
\centering
\includegraphics[width=6.6cm]{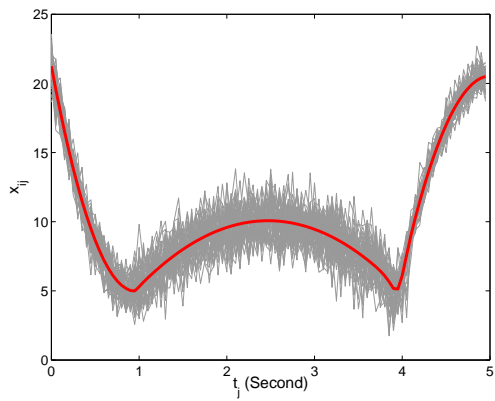}
\caption{Example of $50$ curves simulated according to the proposed regression model with a curves size $m=100$, for the second and the third experiment.}
\label{fig. example of simulated curves of experiment 2 and experiment 3}
\end{figure}

For the second and the third experiments, the curves were simulated with the proposed regression model with hidden logistic process given by equation (\ref{eq.regression model hidden logistic process}). The simulated curves consisted of 3 polynomial regimes ($K=3$) with a polynomial of order $p=2$ with transition time points at $1$ and $4$ seconds. Table \ref{table. parameters of simulation of experiment 2 and experiment 3} shows the set of simulation parameters for these experiments and Fig. \ref{fig. example of simulated curves of experiment 2 and experiment 3} shows an example of simulated curves with $n=50$ and $m=100$. For all the experiments, we considered that the curves were observed over $5$ seconds with a constant sampling period ($\Delta t=t_j-t_{j-1}$ is constant).
\begin{table}[!h]
\centering
\small
\begin{tabular}{|lll|}
\hline
$\bsbeta_1=[23, -36, 18]$& $\bsw_1=[92.72, -46.72]$ &$\sigma_1=1$\\
$\bsbeta_2=[-3.9, 11.08, -2.2]$ & $\bsw_2=[61.16, -15.28]$ &$\sigma_2=1.25$\\
$\bsbeta_3=[-337, 141.5, -14]$ & $\bsw_3=[0,0]$ &$\sigma_3=0.75$ \\
 \hline
\end{tabular}
\caption{Simulation parameters for experiment 2 and experiment 3.}
\label{table. parameters of simulation of experiment 2 and experiment 3}
\end{table}

errer

\begin{figure}[!h]
\centering
\includegraphics[width = 9.5cm,height=7.5cm]{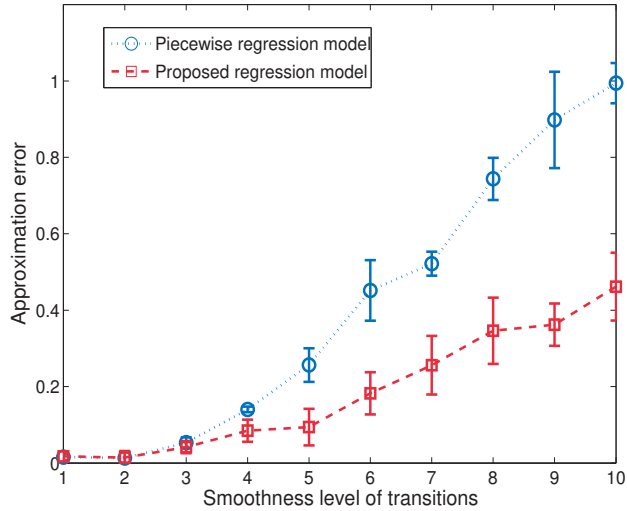}
\caption{Approximation error in relation to the smoothness level of transitions, obtained with the proposed approach (square) and the piecewise polynomial regression approach (circle).}
\label{fig. results_varying_smoothness_level}
\end{figure}

For each value of $n$, each value of $m$ and each value of the smoothness level of transitions, the values of assessment criteria were averaged over $20$ different curves samples. Fig. \ref{fig. results_varying_smoothness_level} shows the error of curves modeling (approximation error) in relation to the smoothness level of transitions. It can be seen that, for abrupt transitions (levels 1, 2 and 3), the two approaches provides similar results. However, when the curves present smooth transitions, the proposed approach provides more accurate results than the piecewise regression approach.

\begin{figure}[!h]
\centering
\includegraphics[width = 9.5cm,height = 7.5cm]{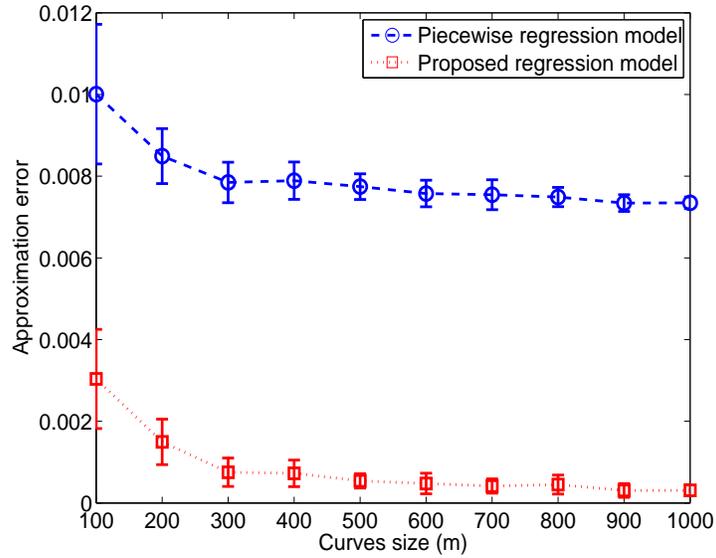}
\caption{Approximation error in relation to the curves size $m$ for $n=50$ curves, obtained with the proposed approach (square) and the piecewise polynomial regression approach (circle).}
\label{fig. results_varying_m}
\end{figure}

\begin{figure}[!h]
\centering
\includegraphics[height = 7cm]{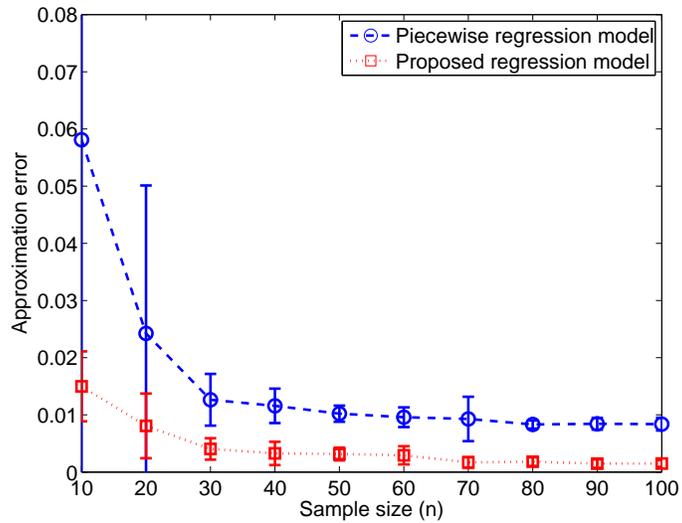}
\caption{Approximation error in relation to the number of curves $n$ for a curves size $m=100$, obtained with the proposed approach (square) and the piecewise polynomial regression approach (circle).}
\label{fig. results_varying_n}
\end{figure}

For the second and the third experiments, it can be seen in both Fig. \ref{fig. results_varying_m} and Fig. \ref{fig. results_varying_n} that the curves modeling error decreases when the curves size $m$ and the number of curves $n$ increase for the two approaches. The results provided by the proposed model are more accurate than those of the piecewise regression approach.

\begin{figure*}[!h]
\centering
\includegraphics[width=6.8cm]{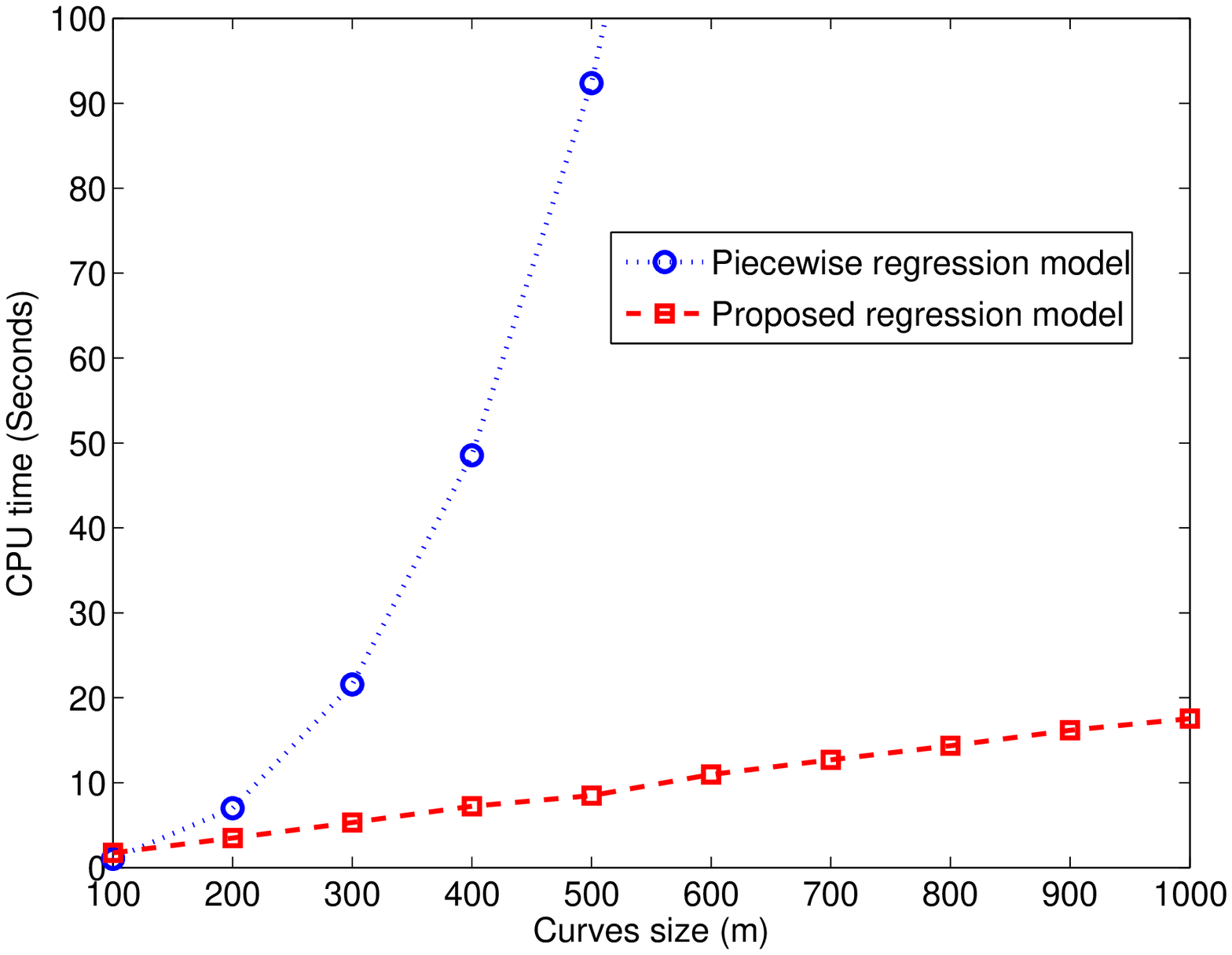}\includegraphics[width=6.8cm]{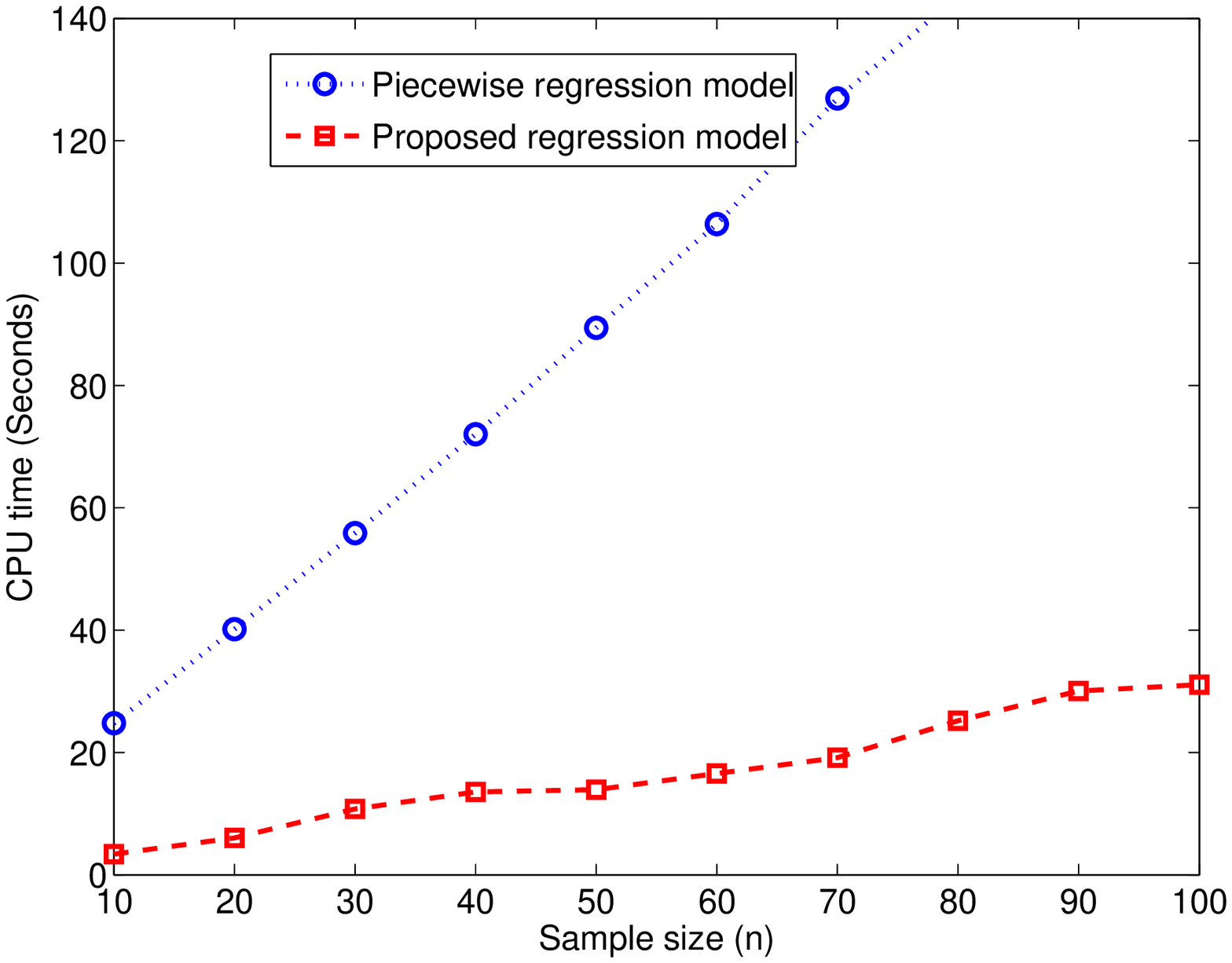}\\
\caption{Average running time (in second) in relation to the curves size $m$ for $n=50$ curves (left) and in relation to the number of curves $n$ for a curves size $m=500$ (right), obtained with the proposed approach (square) and the piecewise polynomial regression approach (circle).}
\label{fig. cputime_varying_m_and_n}
\end{figure*}

Finally, Fig. \ref{fig. cputime_varying_m_and_n} shows that the computation time of the proposed algorithm does not increase much, while that of the piecewise approach grows considerably with the number of curves and with the curves size.

\subsection{Evaluation in terms of curve classification}
\label{ssec: valuation in terms of curve classification}

This section is concerned with the evaluation of the proposed approach in terms of curve classification. Two types of data sets are considered: the waveform data set of Breiman and a real-world data set from railway switch operations.

\subsubsection{Waveform curves of Breiman}
\label{sssec: expermients using waveform data}

In this part, the proposed approach is evaluated in terms of curve classification by considering the waveform data introduced in \cite{breiman} and studied in \cite{hastieANDtibshiraniMDA,RossiConanGuez05NeuralNetworks} and elsewhere. The  waveform data consist in a three-class problem where each curve is generated as follows:
\begin{itemize}
\item $\bx_1(t)=uh_1(t) + (1-u)h_2(t) + \epsilon_t$ for the class 1;
\item $\bx_2(t)=uh_2(t) + (1-u)h_3(t) + \epsilon_t$ for the class 2;
\item $\bx_3(t)=uh_1(t) + (1-u)h_3(t) + \epsilon_t$ for the class 3.
\end{itemize}
where $u$ is a uniform random variable on $(0,1)$,
\begin{itemize}
\item $h_1(t)=\max (6-|t-11|,0)$;
\item $h_2(t)=h_1(t-4)$;
\item $h_3(t)=h_1(t+4)$.
\end{itemize}
and $\epsilon_t$ is a zero-mean Gaussian noise with unit standard deviation.
The temporal interval considered for each curve is $[0;20]$ with a constant period of sampling of 1 second. 500 simulated curves were drawn for each class.

Table \ref{table. classification results waveform data} shows the average classification error rates and the corresponding standard deviations (in parentheses) obtained with the two approaches. It can be observed that the proposed regression approach provides more accurate discrimination results than those of the piecewise polynomial regression approach.

\begin{table}[!h]
\centering
\small
\begin{tabular}{|c|c|}
\hline
Modeling approach & Test error rates (\%)\\
\hline
Piecewise regression model & 2.4 (0.64) \\
Proposed regression model &\textbf{1.67 (0.84)}\\
\hline
\end{tabular}
\caption{Classification results for the waveform curves.}
\label{table. classification results waveform data}
\end{table}

\begin{figure*}[!h]
 \centering
  \includegraphics[height=3.8cm,width=4.48cm]{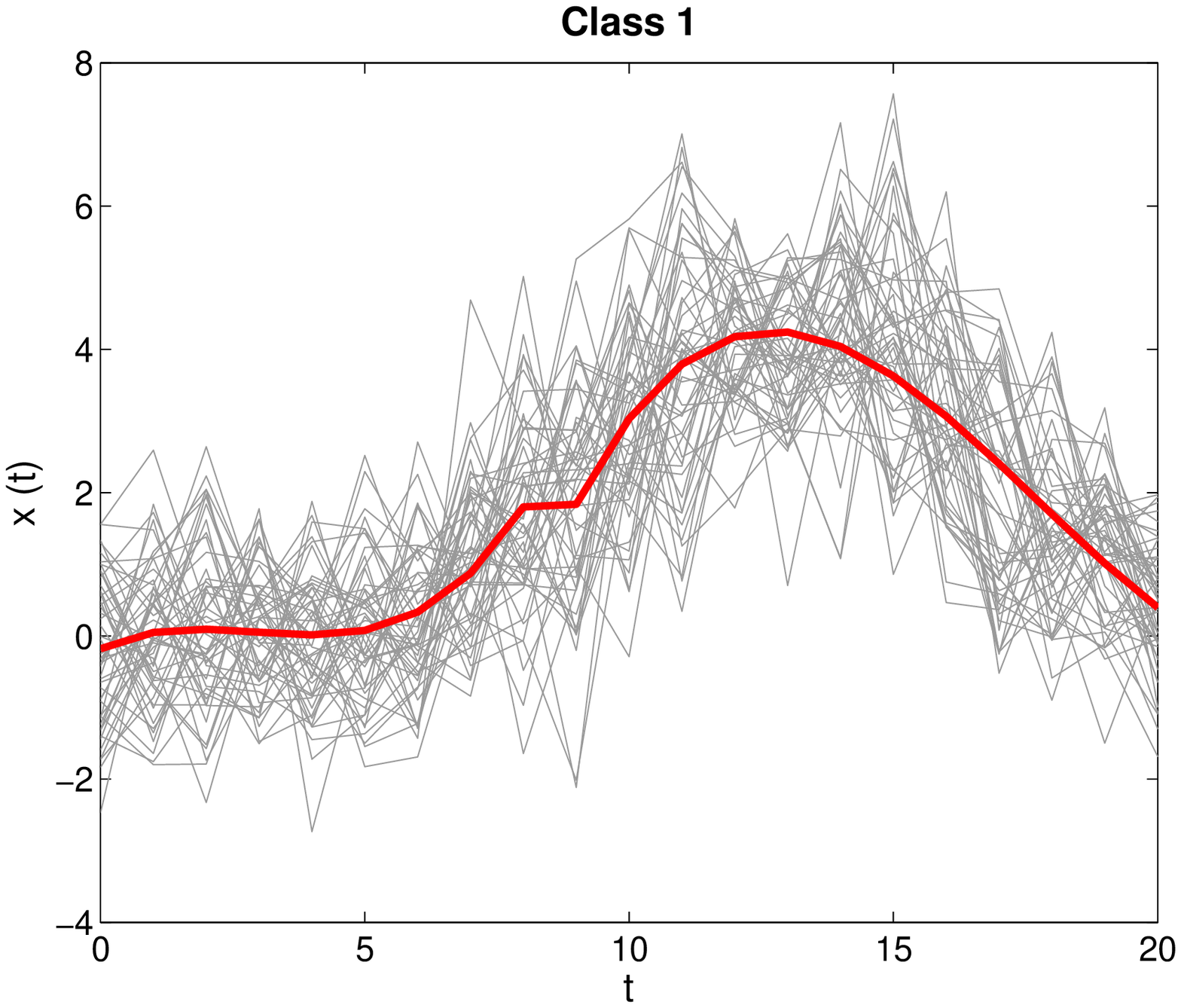}
  \includegraphics[height=3.8cm,width=4.48cm]{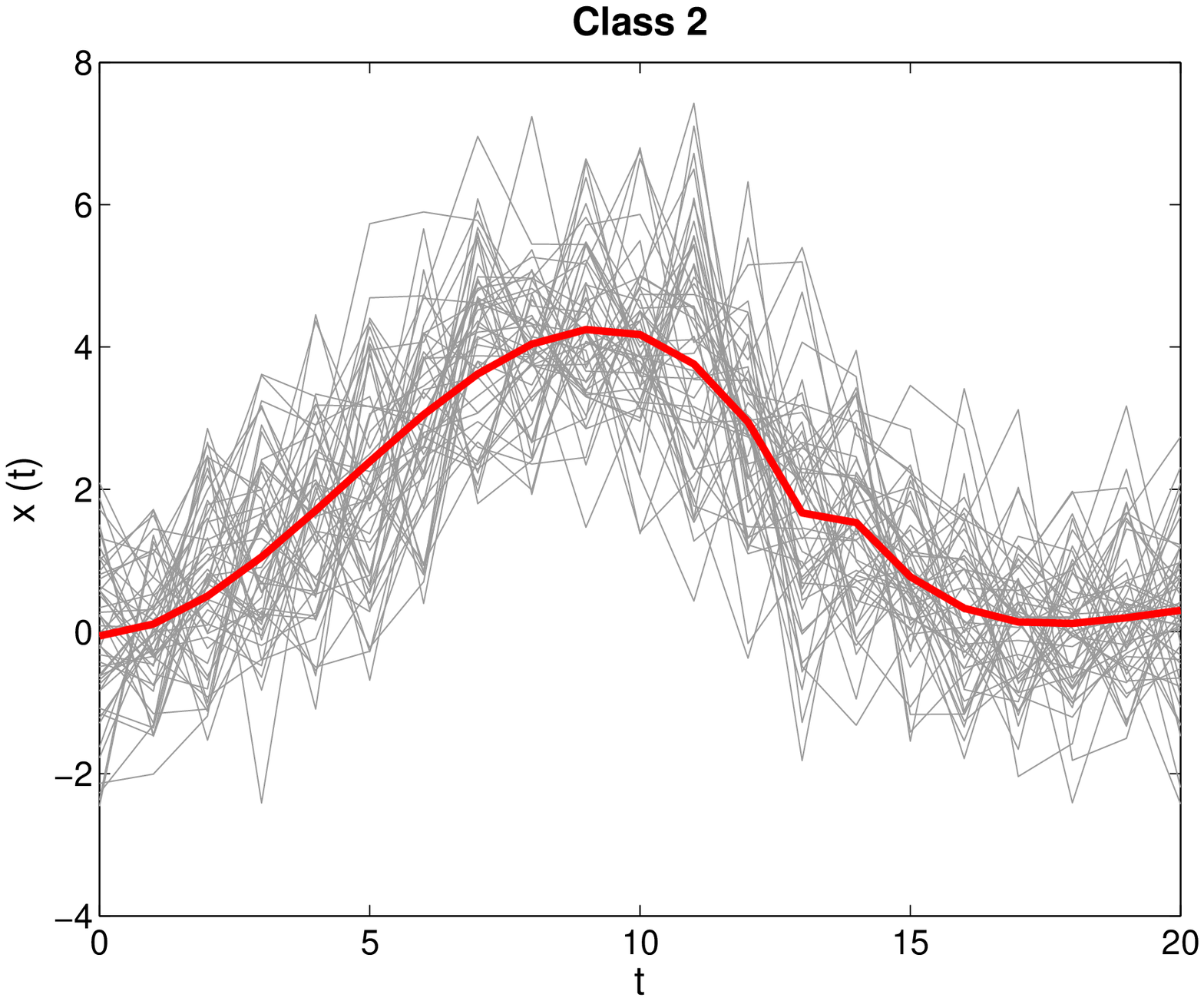}
  \includegraphics[height=3.8cm,width=4.48cm]{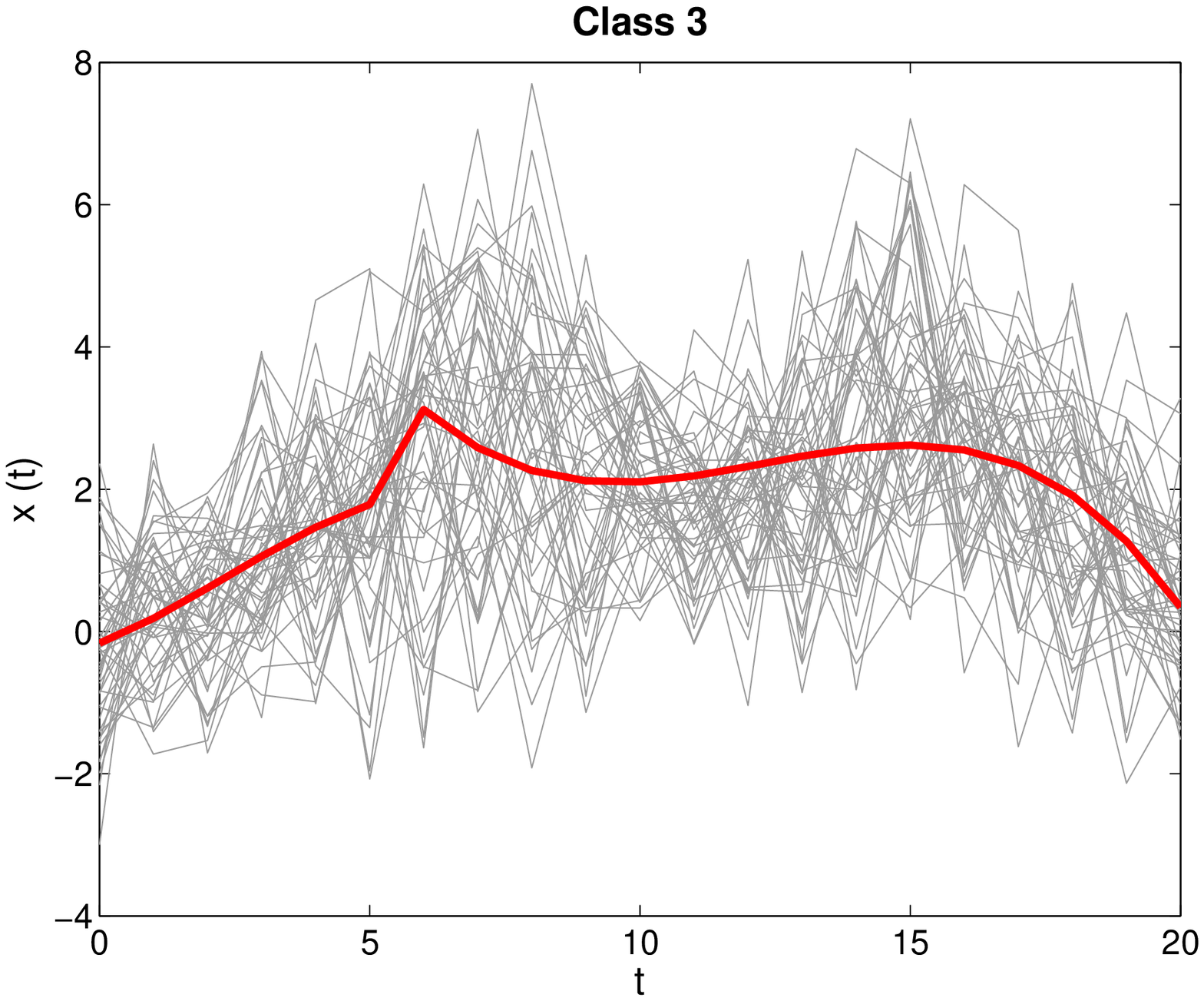}
  \caption{Some examples of the waveforms for the three classes with 50 curves per class and the estimated model for each class obtained with the piecewise polynomial regression approach.}
 \label{fig. waveform_curves_piecewise_model}
\end{figure*}

\begin{figure*}[!h]
 \centering
  \includegraphics[height=3.8cm,width=4.48cm]{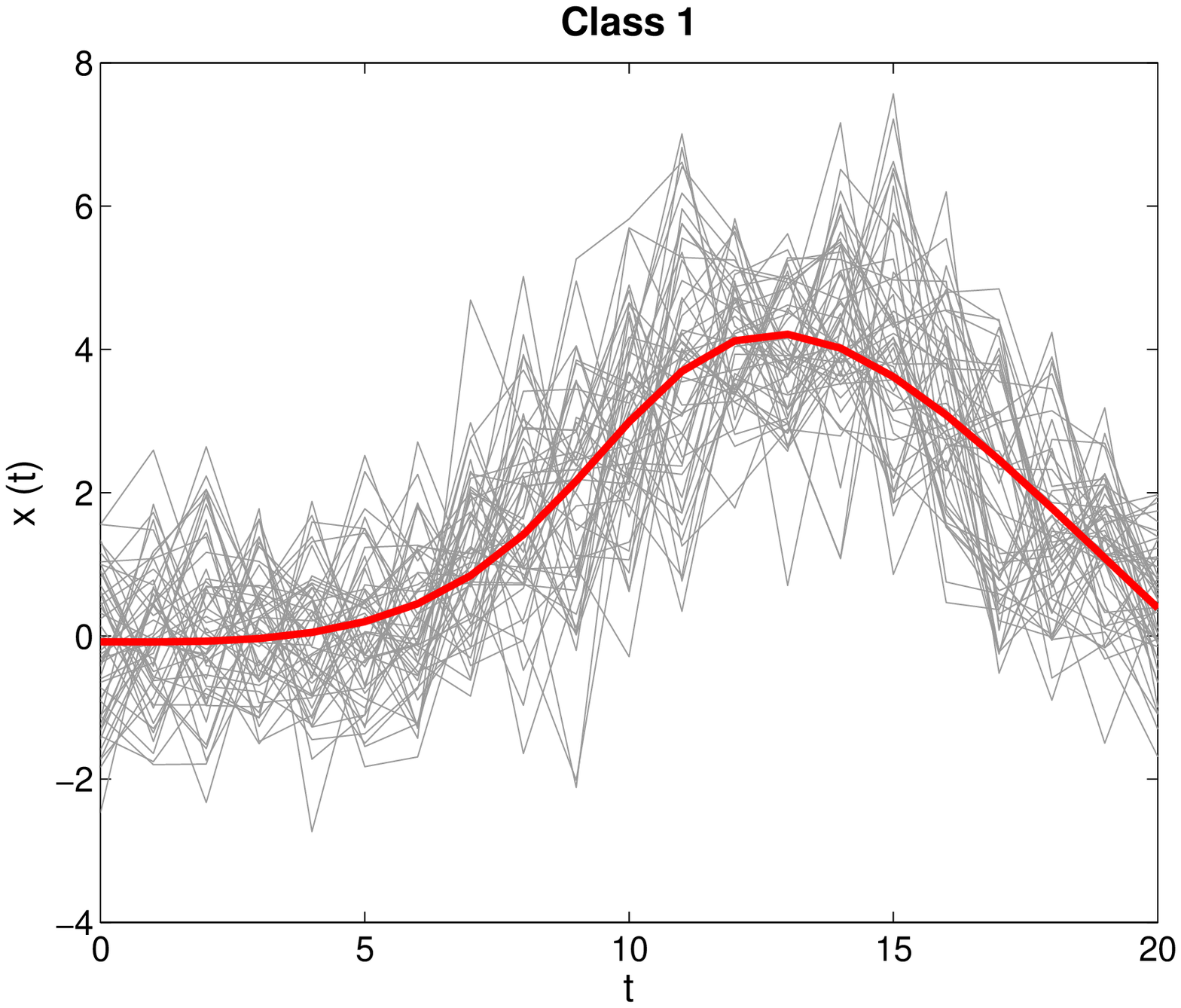}
  \includegraphics[height=3.8cm,width=4.48cm]{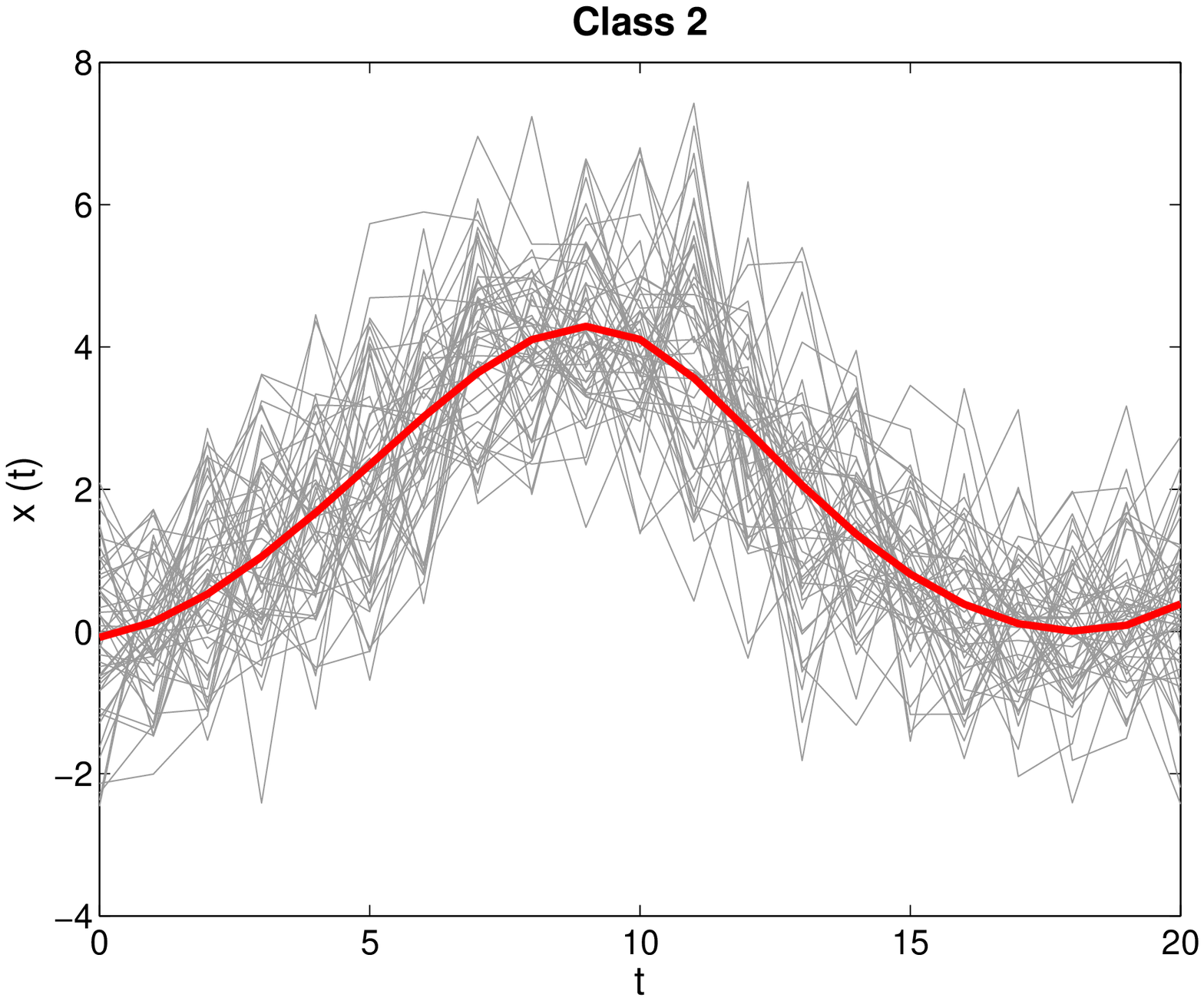}
  \includegraphics[height=3.8cm,width=4.48cm]{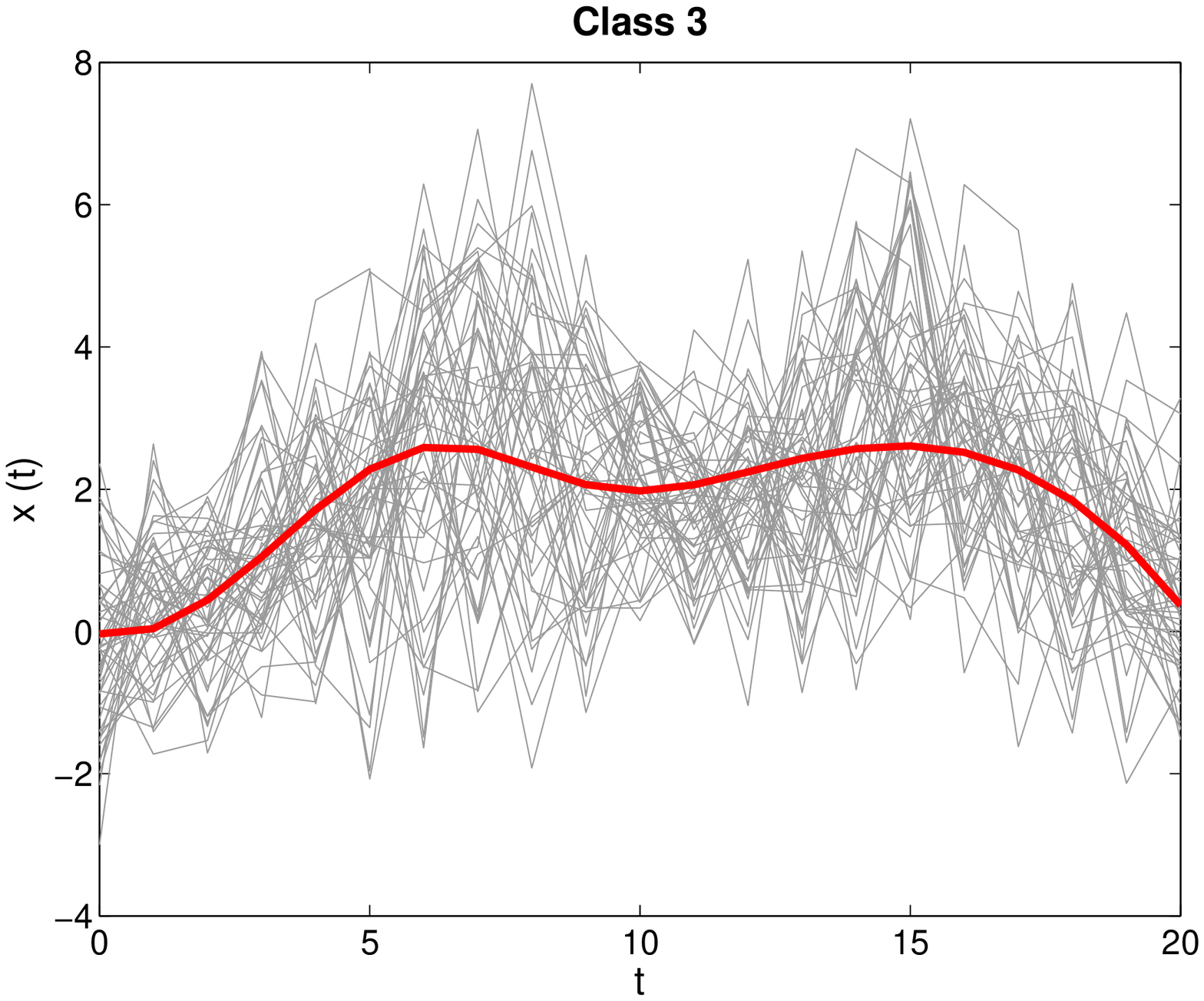}
  \caption{Some examples of the waveforms for the three classes with 50 curves per class and the estimated model for each class obtained with the proposed approach.}
 \label{fig. waveform_curves_proposed_model}
\end{figure*}

Fig. \ref{fig. waveform_curves_piecewise_model} and Fig. \ref{fig. waveform_curves_proposed_model} show the curves estimated respectively by  the piecewise polynomial regression approach and the proposed approach for $K=2$ and $p=3$. We can see that the curve estimated by the piecewise regression approach presents discontinuities since it is computed from a hard segmentation of the curves, while, the curves approximation provided with the proposed regression model is continuous due to the use of the logistic function  adapted to both smooth and abrupt regime changes.


\subsubsection{Real-world curves}
\label{sssec: experiments using real curves}

This section is devoted to an evaluation of proposed approach in terms of curve classification using real curves 
from switch operations. A database of $120$ curves subdivided into three classes was used:
\begin{itemize}
\item $g=1$: no defect class;
\item $g=2$: minor defect class;
\item $g=3$: critical defect class.
\end{itemize}
The cardinal numbers of the three classes are $n_1=35$, $n_2=40$ and  $n_3=45$ respectively. The results in terms of misclassification error rates are given in table \ref{table. classification results for real curves}.
\begin{table}[!h]
\centering
\small
\begin{tabular}{|c|c|}
\hline
Modeling approach & Test error rates (\%)\\
\hline
Piecewise regression model & 1.82 (5.74) \\
Proposed regression model & \textbf{1.67 (2.28)}\\
\hline
\end{tabular}
\caption{Classification results for the switch operation curves.}
\label{table. classification results for real curves}
\end{table}

\begin{figure*}[!h]
\centering
\includegraphics[height=3.5cm,width=4.48cm]{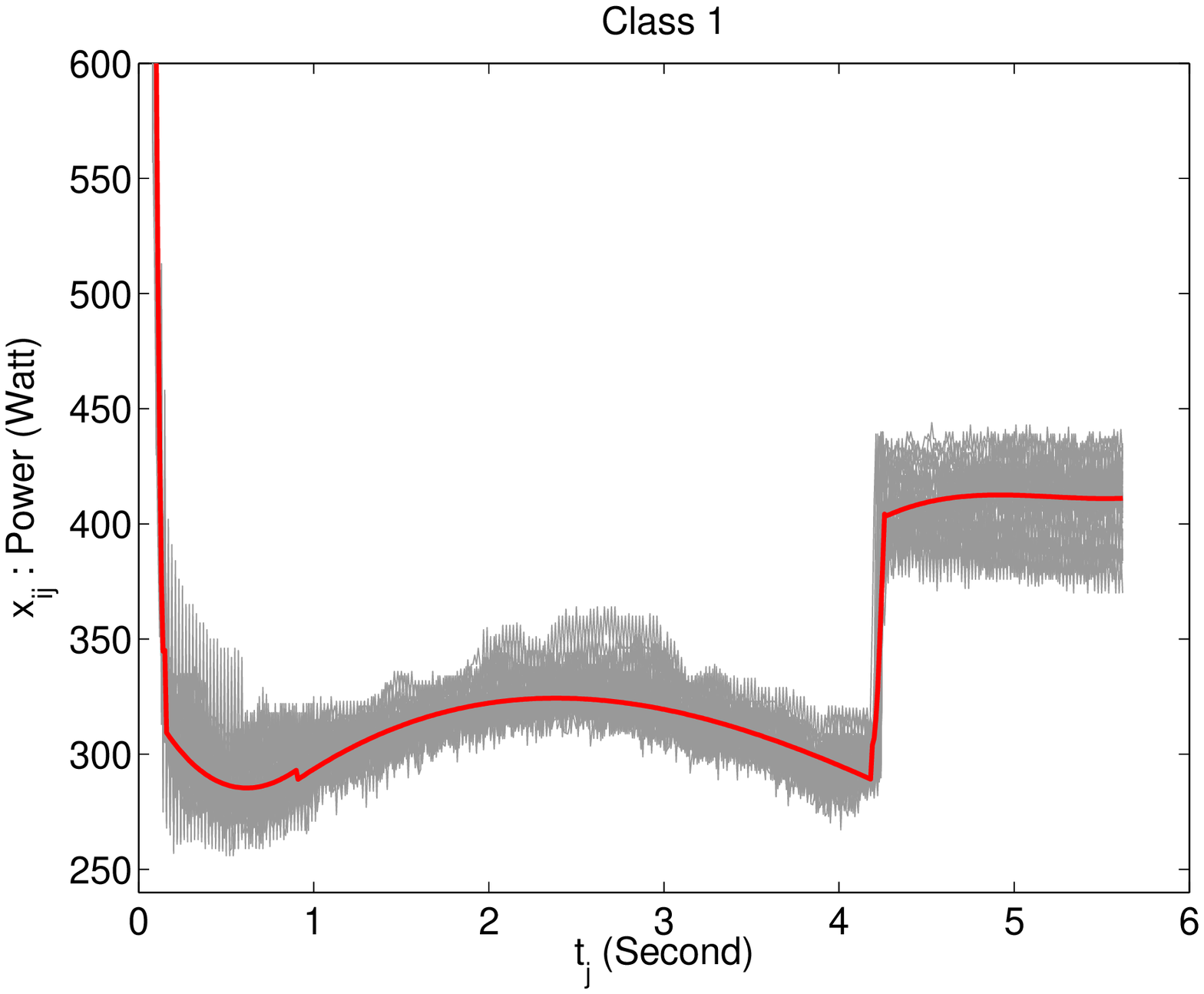}
\includegraphics[height=3.5cm,width=4.48cm]{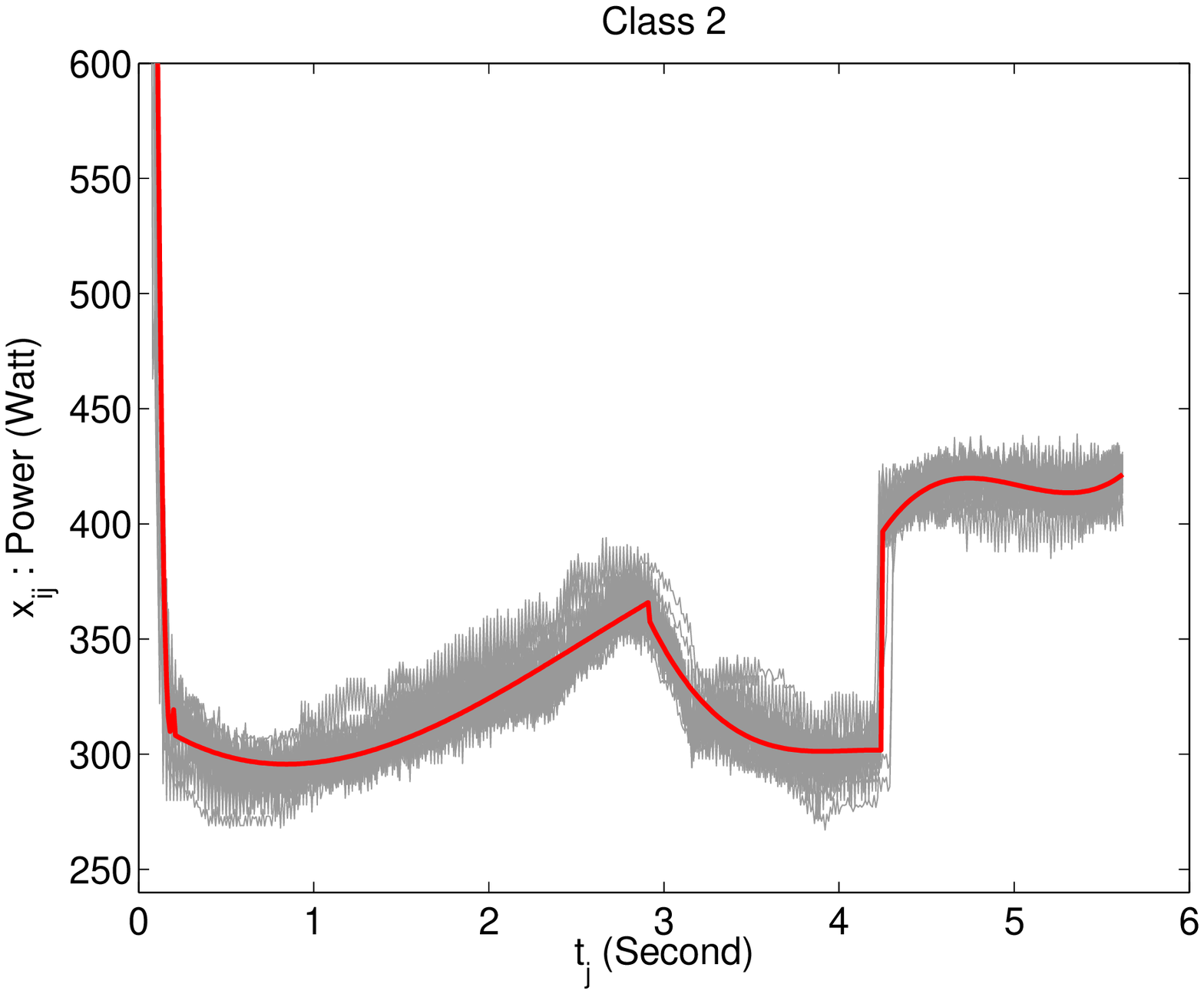}
\includegraphics[height=3.5cm,width=4.48cm]{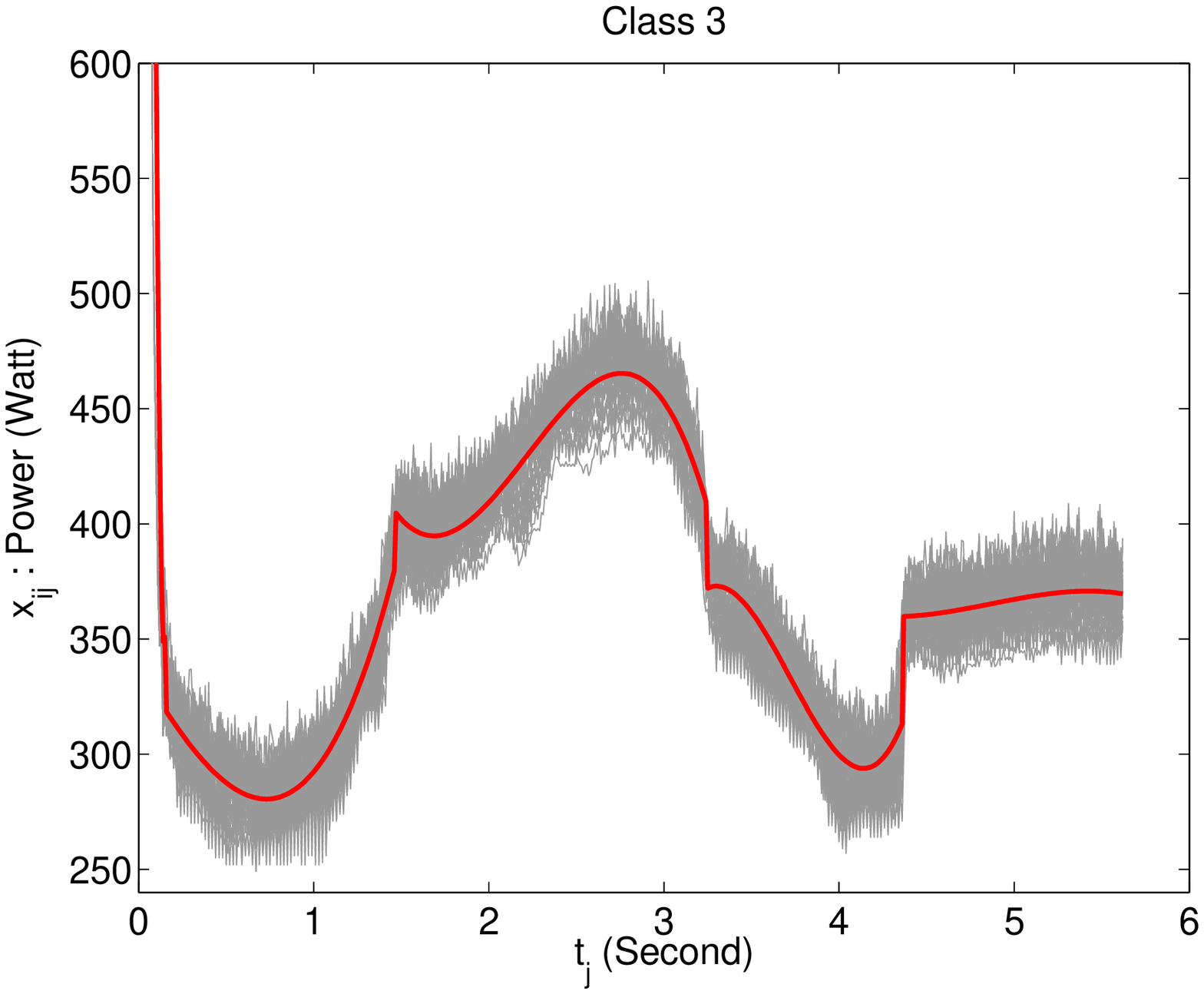}\\
\hspace{0.07 in}\includegraphics[height=3.5cm,width=4.32cm]{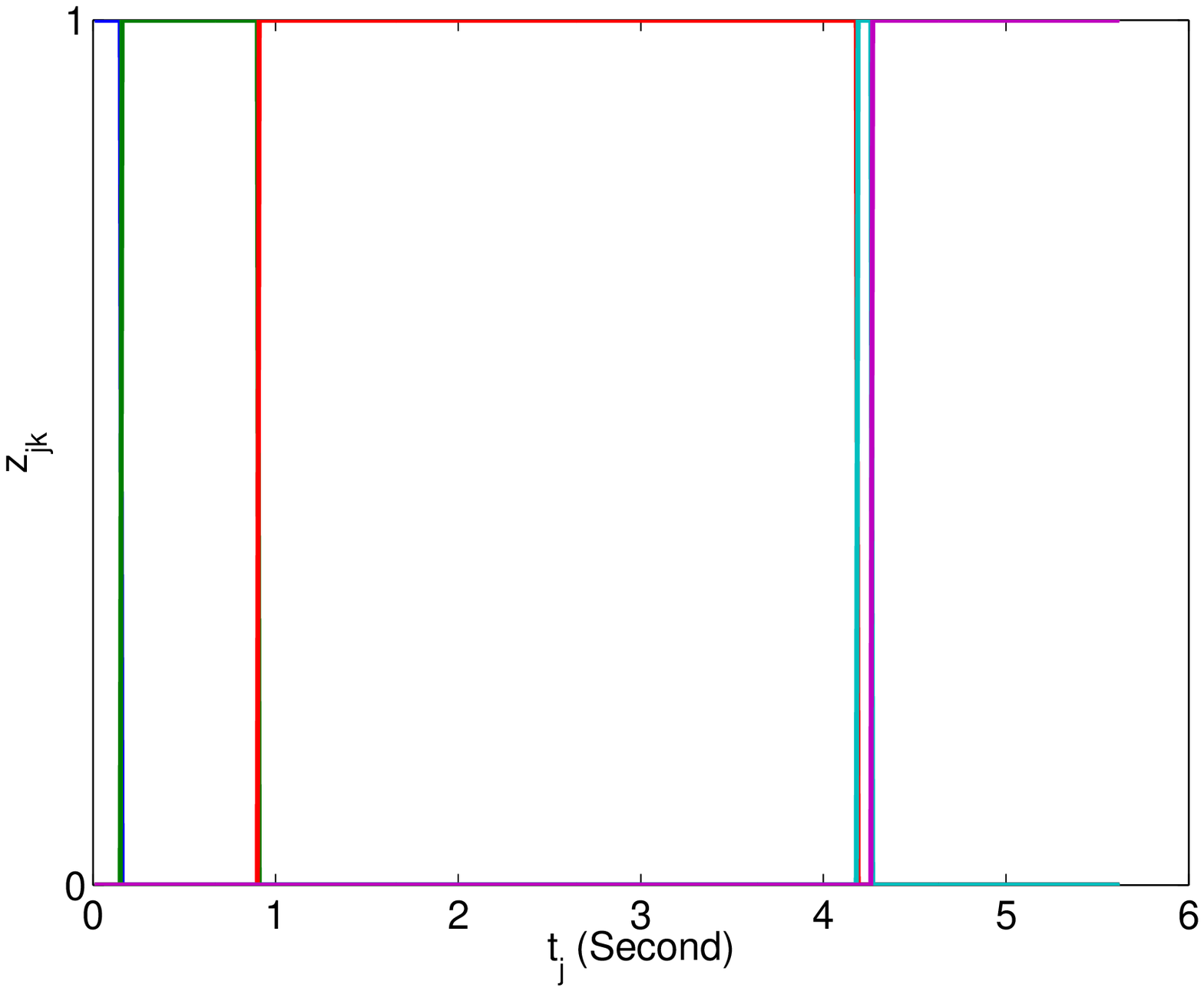}
\hspace{.04 in}\includegraphics[height=3.5cm,width=4.32cm]{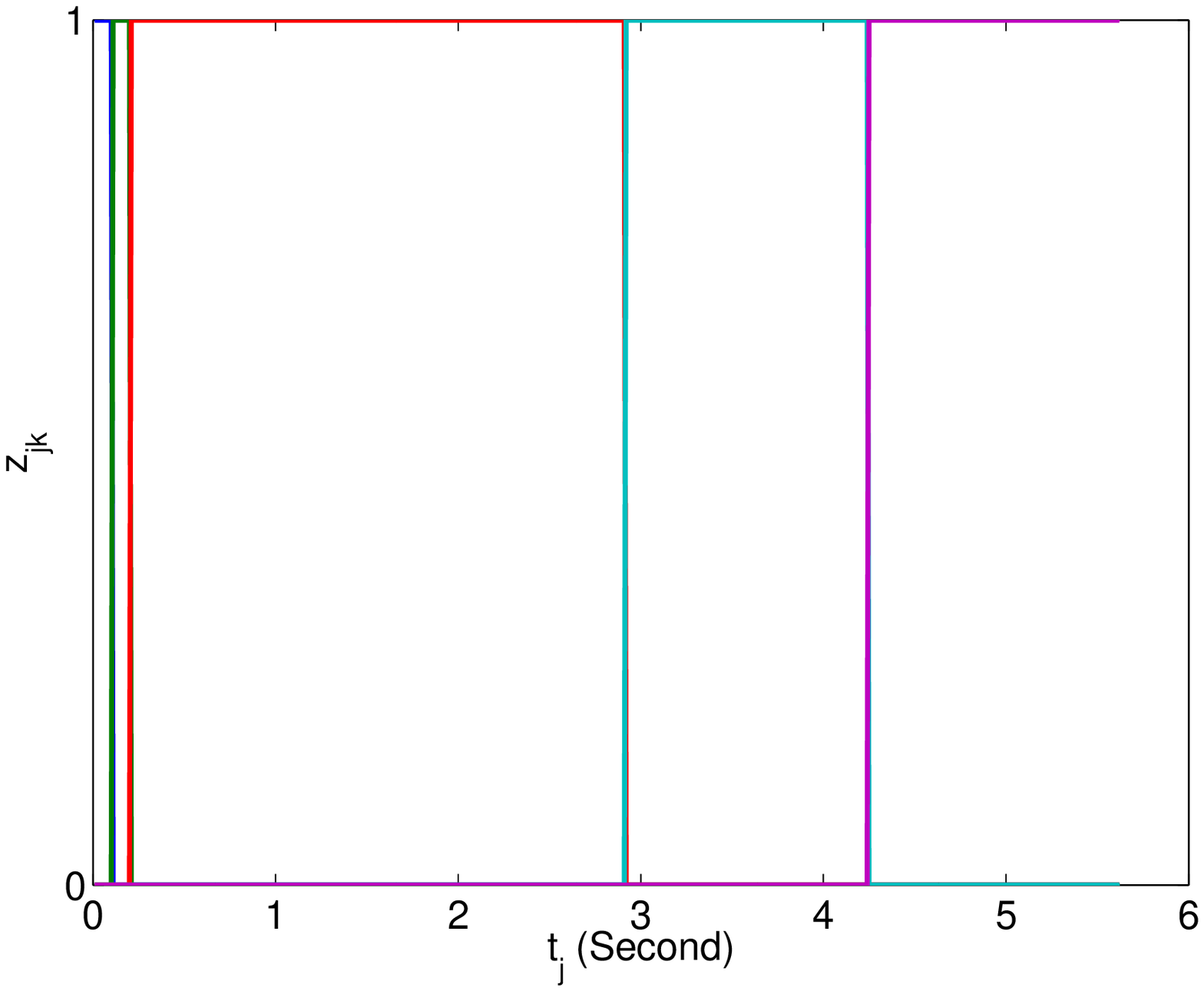}
\hspace{.04 in}  \includegraphics[height=3.5cm,width=4.32cm]{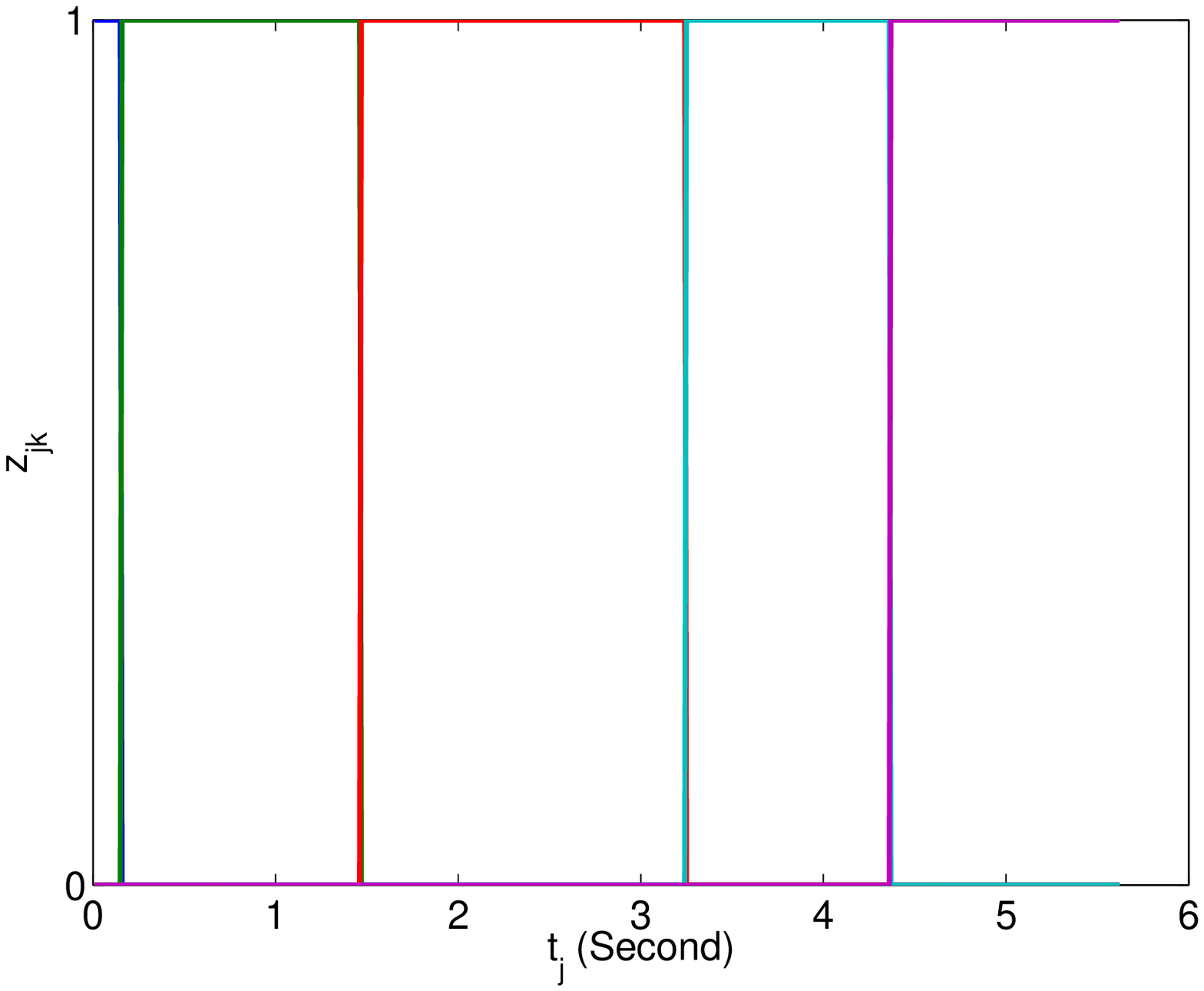}
\caption{The three classes of switch operation curves and the corresponding estimated curve (top) and the segmentation (bottom) obtained with the piecewise polynomial regression model.}
\label{fig. real curves approximations with the piecewise regression approach}
\end{figure*}

\begin{figure*}[!h]
\centering
\includegraphics[height=3.5cm,width=4.48cm]{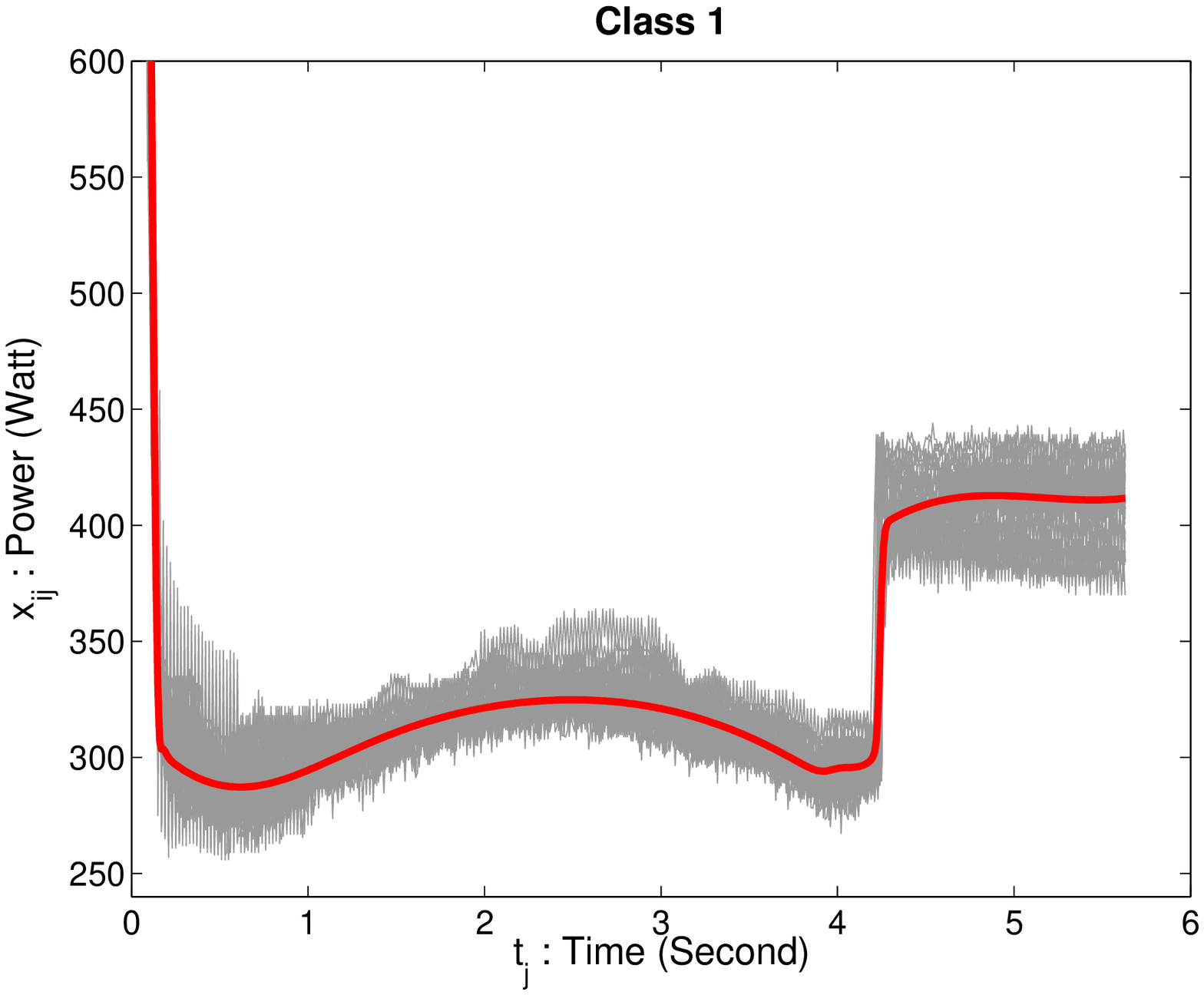}
\includegraphics[height=3.5cm,width=4.48cm]{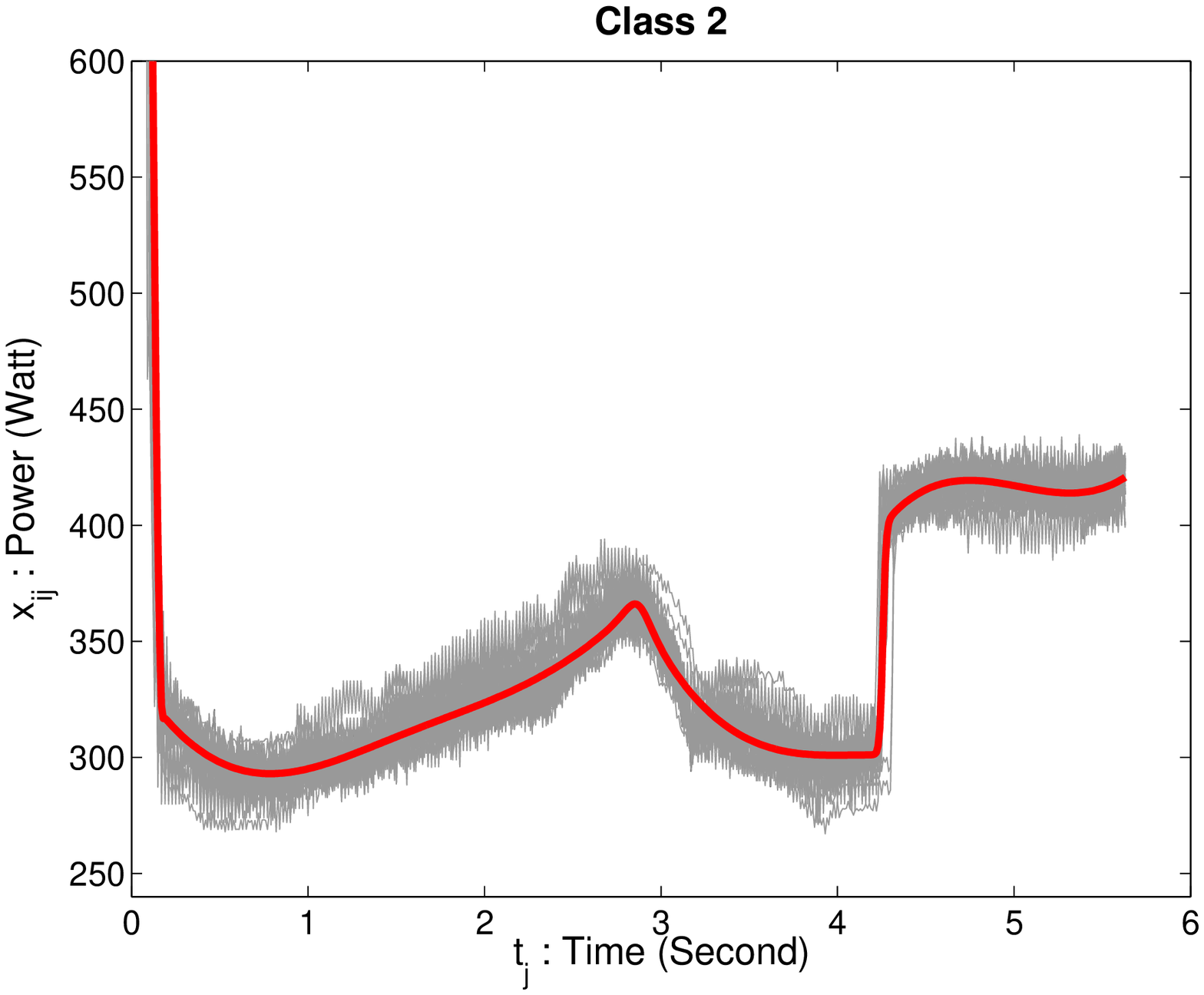}
 \includegraphics[height=3.5cm,width=4.48cm]{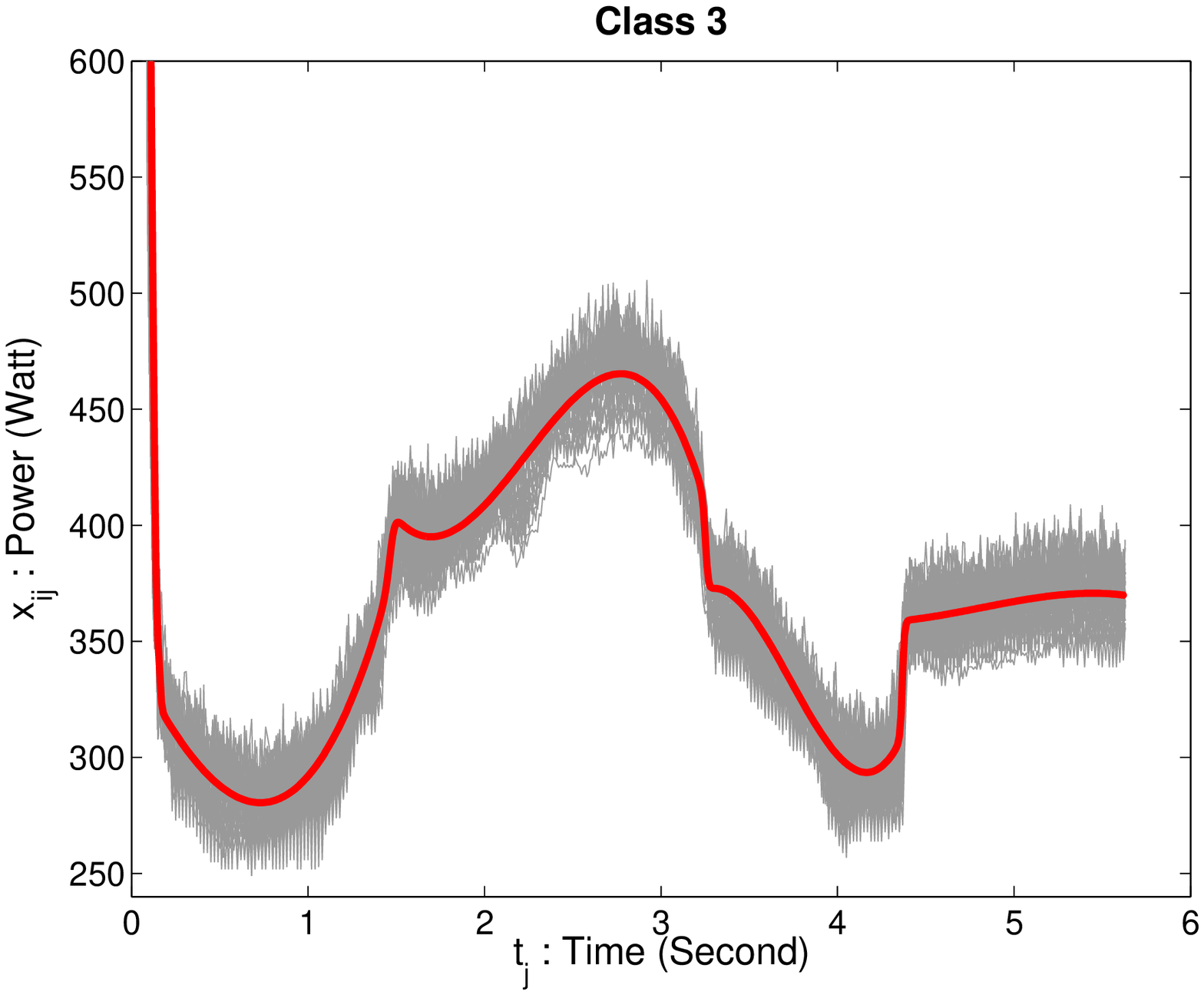} \\
 \includegraphics[height=3.5cm,width=4.48cm]{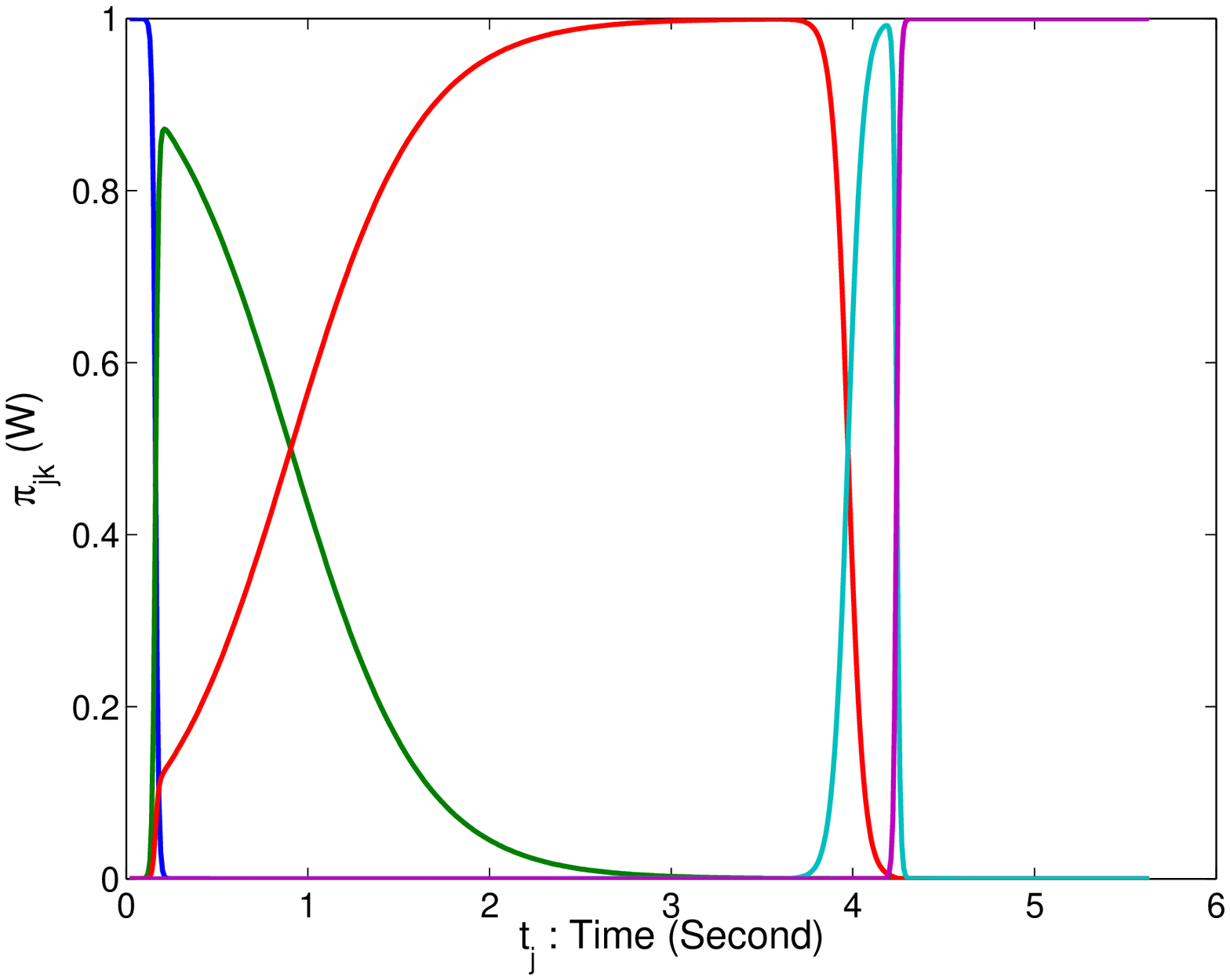}
 \includegraphics[height=3.5cm,width=4.48cm]{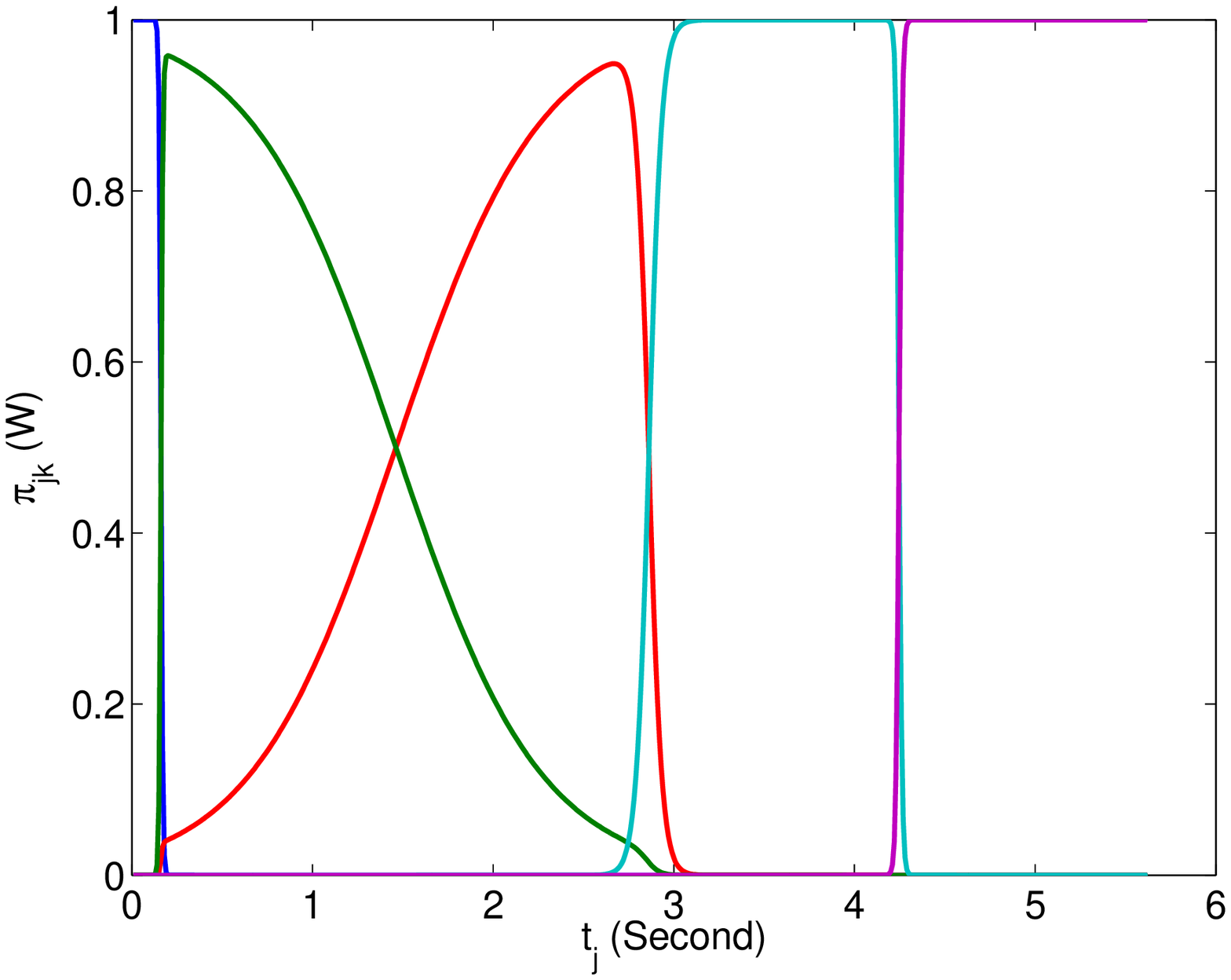}
 \includegraphics[height=3.5cm,width=4.48cm]{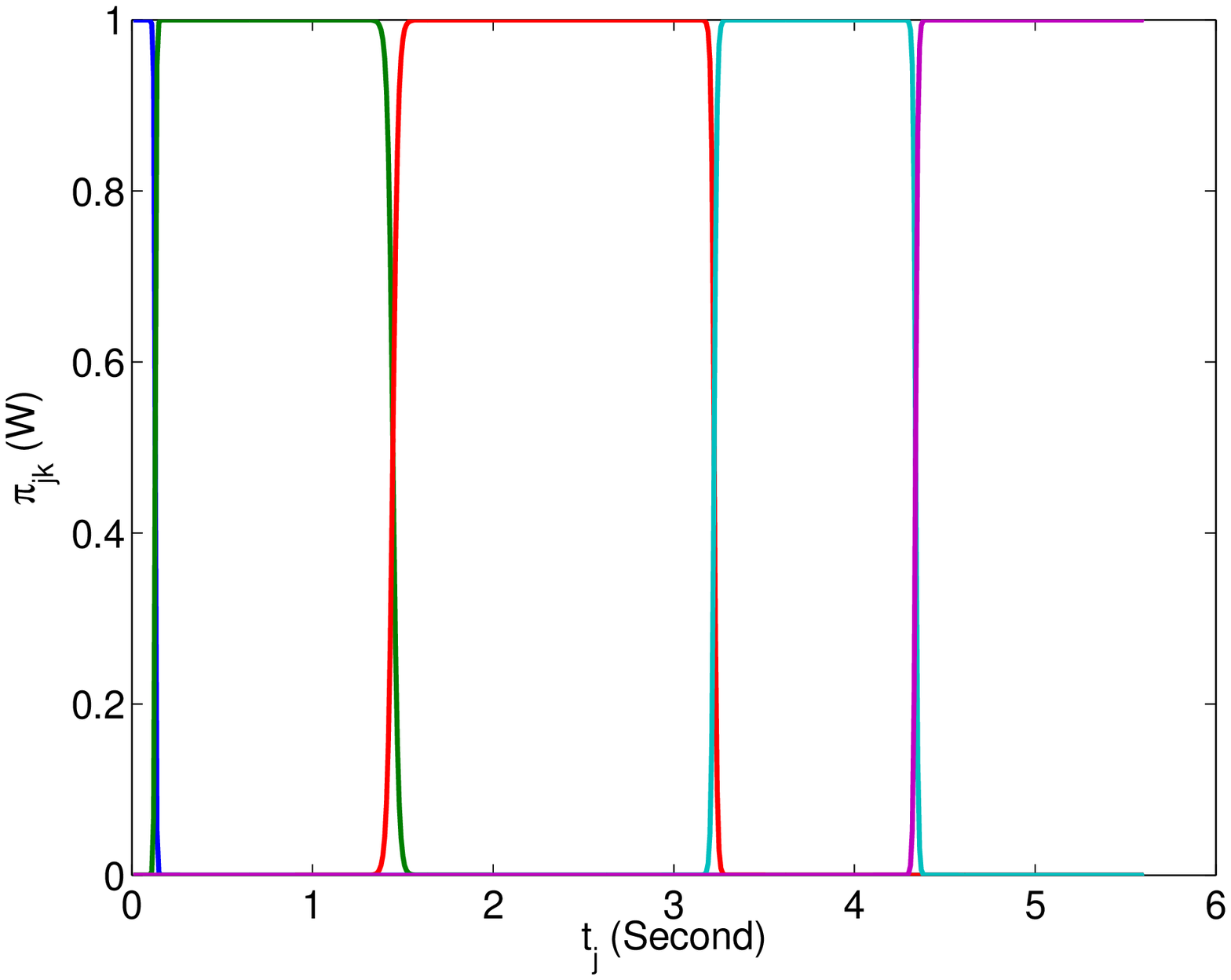}
  \caption{The three classes of the switch operation curves and the estimated curves and the corresponding estimated logistic probabilities for each class of curves obtained provided by the proposed approach.}
\label{fig. curves approximation and process probabilities for the real curves}
\end{figure*}

We can see that the proposed approach provides results slightly better than those of the piecewise regression approach. Although the difference in terms of misclassification error is not very significant, the proposed method ensure, unlike the piecewise regression approach,  the continuity of the estimated curves (see Fig.\ref{fig. real curves approximations with the piecewise regression approach} and Fig. \ref{fig. curves approximation and process probabilities for the real curves}). Since the transitions involved in the segments 1 and 5 are abrupt, they are well estimated by the two approaches. However, the segments 2 and 4 estimated by the proposed approach have been found more realistic regarding the true phases involved in  a real switch operation.

It should be mentioned that, despite the very good results obtained for curves description, this approach may have limitations in terms of curve classification. The next section illustrates this limitation through simulations.

\subsection{Behaviour of the proposed approach for complex shaped classes}

The aim of this part is to show the behaviour of the proposed classification approach for classes having complex non convex shapes. We consider simulated curves from two classes:

\begin{itemize}
\item class 1: this class consists of 40 curves simulated  with the generative model presented in section \ref{ssec: the generative model of curves}. 15 curves are simulated with the parameters estimated from the first class of the real dataset  (see Fig. \ref{fig. curves approximation and process probabilities for the real curves} (left))  and 25 curves are simulated with the parameters estimated from the second class (see Fig. \ref{fig. curves approximation and process probabilities for the real curves} (middle)),

\item class 2: this class consists of 37 curves simulated  in the following manner:  17 curves are simulated with the parameters estimated from the second class of the real data and 20 curves are simulated with the parameters estimated from the third  class (cf. Fig. \ref{fig. curves approximation and process probabilities for the real curves} (middle) and Fig. \ref{fig. curves approximation and process probabilities for the real curves} (right) respectively).
\end{itemize}

\begin{figure*}[!h]
 \centering
 \begin{tabular}{cc}
 \includegraphics[width=4.9cm]{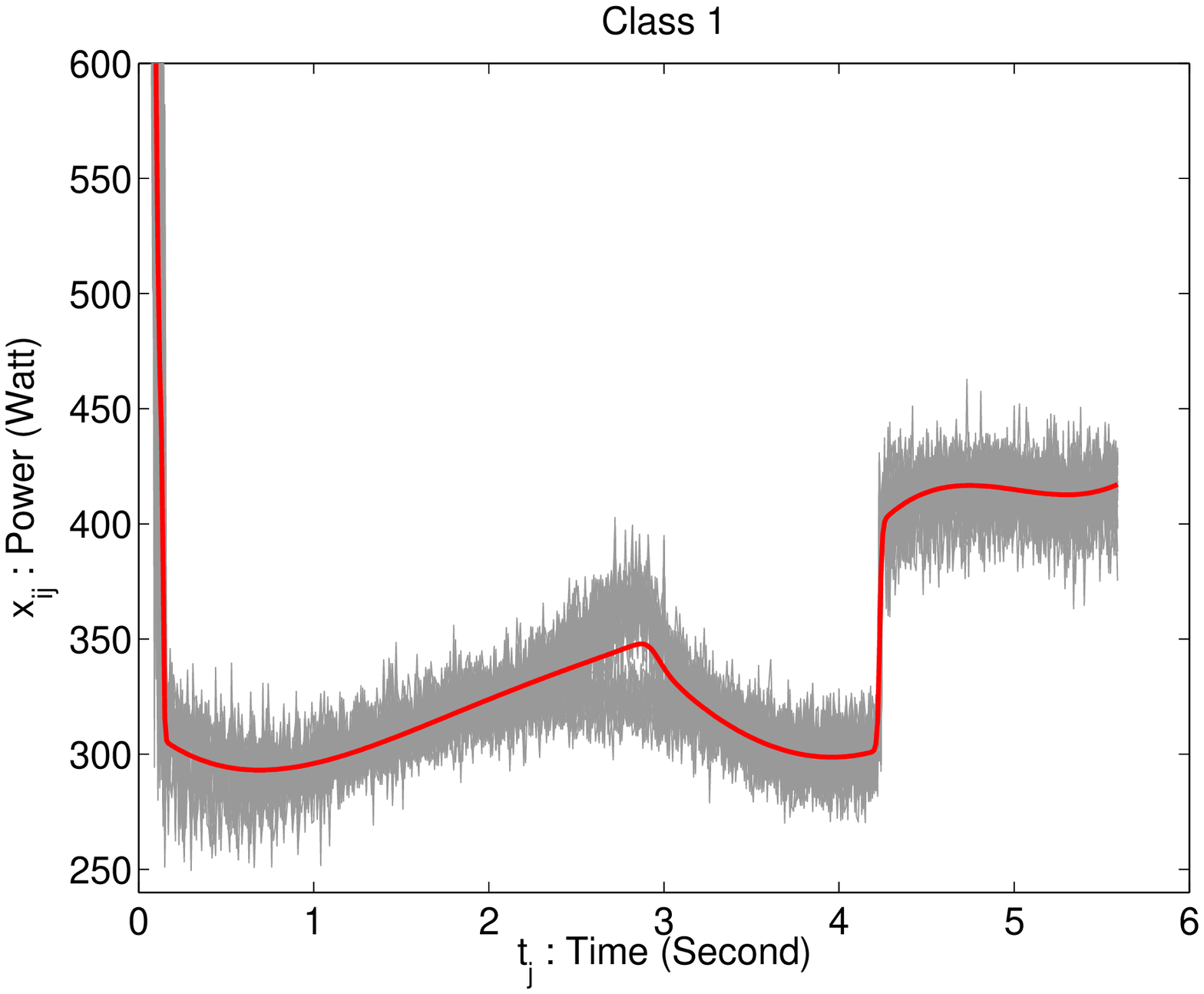}&
 \includegraphics[width=4.9cm]{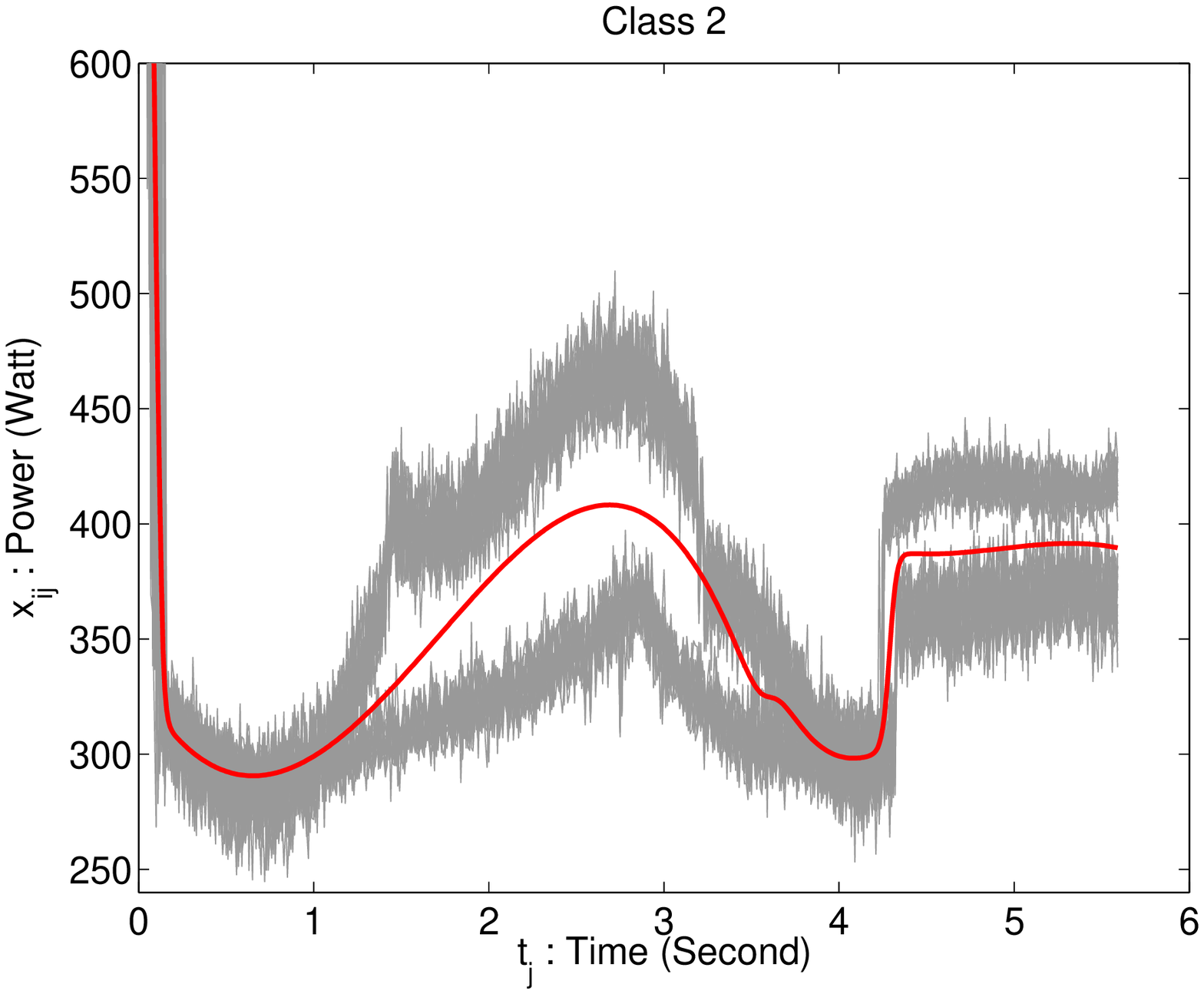}\\
 \includegraphics[width=4.9cm]{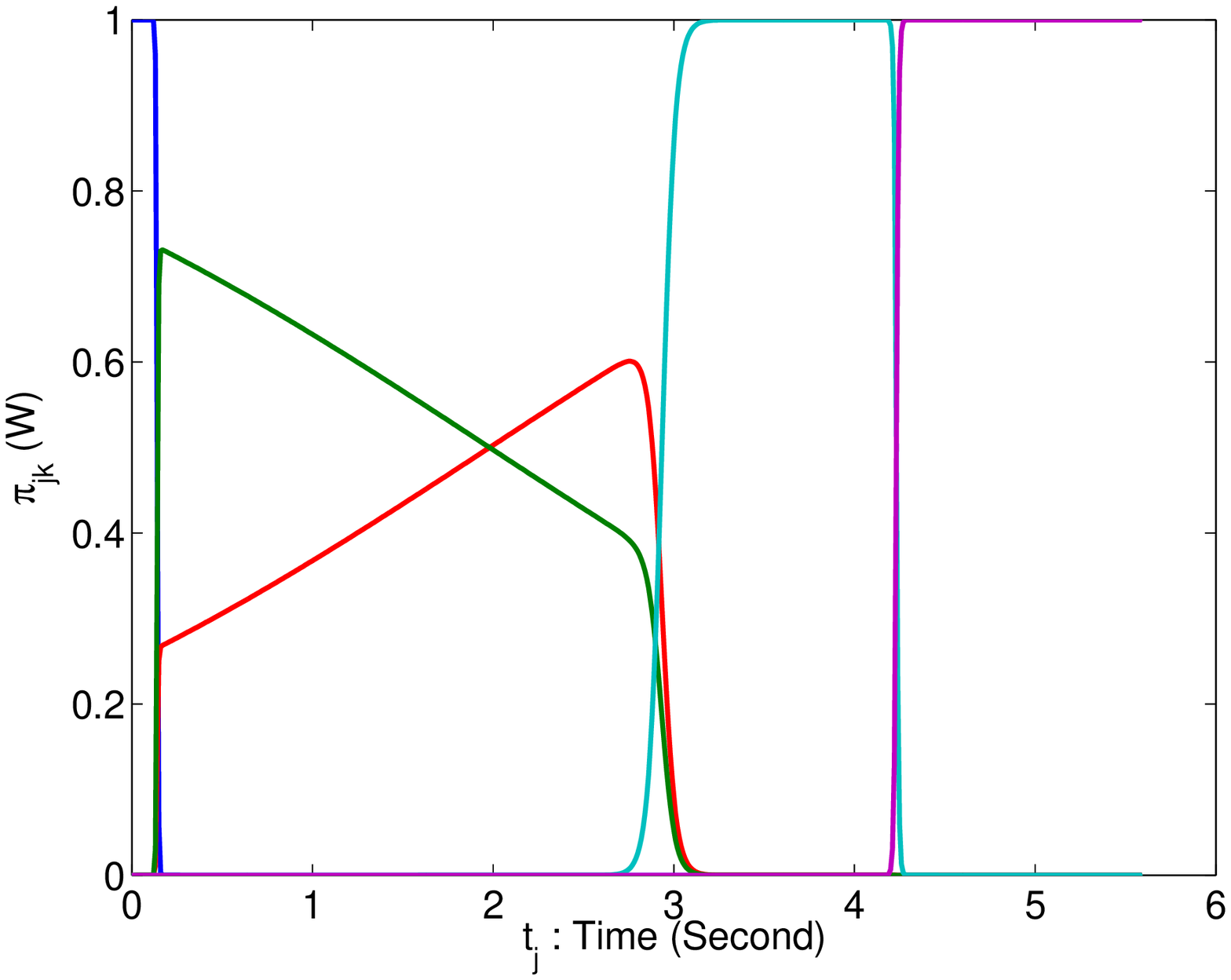}&
 \includegraphics[width=4.9cm]{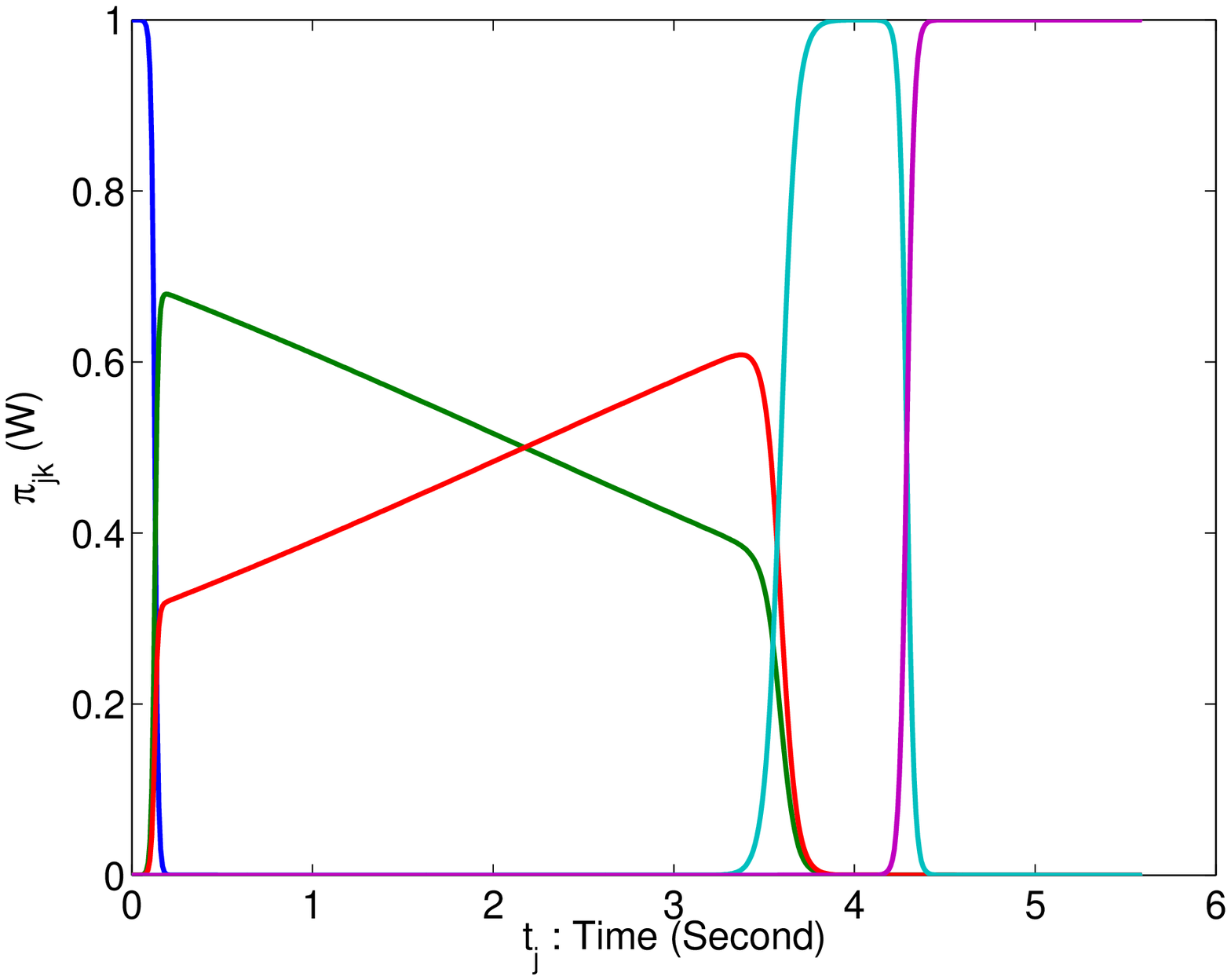}
 \end{tabular}
 \caption{Two complex classes (40 curves from class 1 and 37 curves from class 2) generated according to the generative model presented in section \ref{ssec: the generative model of curves}.}
\label{fig. shape variability illustration with the generative model}
\end{figure*}

Fig. \ref{fig. shape variability illustration with the generative model} shows the  simulated curves, the estimated curves approximation and the estimated logistic probabilities.

In this setting, the classification error rate obtained with the proposed approach is 20 \% with a standard deviation of 8.16\%. The poor performances in this case can be attributed to the non homogeneous nature of the simulated groups. As it can be observed in Fig. \ref{fig. shape variability illustration with the generative model}, especially  for class 2, the proposed model is not adapted for classes having complex shape.

\section{Conclusion}

This paper introduces a new approach for  functional data description. The proposed approach consists in a regression model governed by a discrete hidden process. The logistic functions used as the probability distributions of the hidden variables allow for smooth or abrupt transitions between various polynomial regression components over time. A curves discrimination method is derived by applying the Maximum A Posteriori rule. An experimental study performed on simulated data and real curves acquired during switch operations reveals good performances of the proposed approach in terms of curve modeling and classification,  compared to the piecewise polynomial regression approach. The limitations of the proposed approach in terms of classification have been shown for complex shaped classes. A future work will consist in considering a more efficient approach to deal with these limitations,  where a complex shaped class will be modeled by a mixture of hidden process regression models.

\section*{Appendix: The EM algorithm}

In the context of maximizing the likelihood by the EM algorithm, the complete log-likelihood \cite{dlr} is written as:
\begin{eqnarray}
CL(\bstheta;\mathcal{X},\bz)&=&\log p(\bx_1,\ldots,\bx_n,\bz|\mathbf{x};\bstheta)\nonumber\\
&=&\log \prod_{i=1}^n \prod_{j=1}^m \prod_{k=1}^K \left[p(z_j=k;\bw) p(x_{ij}|z_j=k;\bstheta)\right]^{z_{jk}} \nonumber\\
&=&\sum_{i=1}^{n}\sum_{j=1}^{m} \sum_{k=1}^K z_{jk}\log
[\pi_{jk}(\bw)\mathcal{N}\left(x_{ij};\bsbeta^T_k\bsr_{j},\sigma^2_k\right)].
\label{eq. complete log-likelihood}
\end{eqnarray}
The EM algorithm starts from an initial parameter $\bstheta^{(0)}$ and alternates the two following steps until convergence:

\subsubsection*{\textbf{E Step (Expectation)} }

This step computes the conditional expectation of the complete log-likelihood given the observations and the current value $\bstheta^{(q)}$: 
\begin{eqnarray}
Q(\bstheta,\bstheta^{(q)})&\!\!\!\!=\!\!\!\!& E\left[CL(\bstheta;\mathcal{X},\bz)|\mathcal{X};\bstheta^{(q)}\right]\nonumber\\
&\!\!\!\!=\!\!\!\!& \sum_{i=1}^{n}\sum_{j=1}^{m}\sum_{k=1}^K E (z_{jk}|x_{ij};\bstheta^{(q)}) \log \left[\pi_{jk}(\bw)\mathcal{N}(x_{ij};\bsbeta^T_k\bsr_{j},\sigma_k^2)\right] \nonumber \\
&\!\!\!\!=\!\!\!\! &\sum_{i=1}^{n}\sum_{j=1}^{m}\sum_{k=1}^K \tau^{(q)}_{ijk}\log \left[\pi_{jk}(\bw)\mathcal{N} \left(x_{ij};\bsbeta^T_k\bsr_{j},\sigma^2_k \right)\right] \nonumber \\
&\!\!\!\!=\!\!\!\! &\sum_{i=1}^{n}\sum_{j=1}^{m}\sum_{k=1}^K \tau^{(q)}_{ijk}\log \pi_{jk}(\bw) \! + \! \sum_{i=1}^{n}\sum_{j=1}^{m}\sum_{k=1}^K \! \tau^{(q)}_{ijk}\log \mathcal{N} \left(x_{ij};\bsbeta^T_k\bsr_{j},\sigma^2_k \right),\nonumber
\end{eqnarray}
where $\tau^{(q)}_{ijk}$ is the posterior probability that $x_{ij}$ originates from the $k$th regression model defined by equation (\ref{eq.tik}).
\\As shown in the expression of $Q$, this step simply requires the computation of $\tau^{(q)}_{ijk}$.

\subsubsection*{\textbf{M step (Maximization)} }

This step updates the value of the parameter $\bstheta$ by maximizing $Q$ with respect to $\bstheta$. To perform this maximization, we can see that $Q$ is written as: 
\begin{equation}
Q(\bstheta,\bstheta^{(q)})=Q_1(\bw)+\sum_{k=1}^K Q_2(\bsbeta_k,\sigma^2_k),
\end{equation}
with
\begin{equation}
Q_1(\bw)=\sum_{i=1}^{n}\sum_{j=1}^{m}\sum_{k=1}^K \tau^{(q)}_{ijk}\log \pi_{jk}(\bw)
\end{equation}
and
\begin{eqnarray}
Q_2(\bsbeta_k,\sigma^2_k) &=& \sum_{i=1}^{n} \sum_{j=1}^{m}\tau^{(q)}_{ijk}\log \mathcal{N}\left(x_{ij};\bsbeta^T_k\bsr_{i},\sigma^2_k\right) \nonumber \\
&=& -\frac{1}{2} \left[\frac{1}{\sigma_k^2}\sum_{i=1}^{n}\sum_{j=1}^{m}\tau^{(q)}_{ijk}
\left(x_{ij}-\bsbeta^T_k\bsr_{j}\right)^2 +n m_k^{(q)}\log \sigma_k^2 \right] \nonumber \\
& &-\frac{n m_k^{(q)}}{2}\log 2\pi \quad ; \quad k=1,\ldots,K ,
\end{eqnarray}
where $m_k^{(q)}=\sum_{j=1}^{m} \tau^{(q)}_{ijk}$ is the number of elements in component $k$ estimated at iteration $q$ for each curve $\bx_i$. Thus, the maximization of $Q$ can be performed by separately maximizing $Q_1(\bw)$ with respect to $\bw$ and $Q_2(\bsbeta_k, \sigma^2_k)$ with respect to $(\bsbeta_k,\sigma^2_k)$ for $k=1,\ldots,K$.

Maximizing  $Q_2$ with respect to $\bsbeta_k$ consists in analytically solving a weighted least-squares problem. The estimates are given by:
\begin{equation}
\bsbeta_k^{(q+1)}  = \arg \min \limits_{\substack {\bsbeta_k}} \sum_{i=1}^{n}\sum_{j=1}^{m} \tau^{(q)}_{ijk} ( x_{ij}-\bsbeta_k^{T}\bsr_j)^2 = (\Lambda^T W_k^{(q)}\Lambda)^{-1}\Lambda^T W_k^{(q)} \mathcal{X}.
\end{equation} 

Maximizing  $Q_2$ with respect to $\sigma_k^2$ provides the following updating formula:
\begin{eqnarray}
{\sigma}_k^{2(q+1)}&=& \arg \min \limits_{\substack{\sigma_k^2}} \left[\frac{1}{\sigma_k^2}\sum_{i=1}^{n}\sum_{j=1}^{m} \tau^{(q)}_{ijk}  (x_{ij}-\bsbeta^{T(q+1)}_k\bsr_{j} )^2 +n m_k^{(q)}\log \sigma_k^2 \right] \nonumber \\
&=& \frac{1}{n m_k^{(q)}}\sum_{i=1}^{n}\sum_{j=1}^{m} \tau^{(q)}_{ijk} (x_{ij}-{\bsbeta}_k^{T(q+1)}\bsr_j)^2 .
\end{eqnarray}
\section*{Acknowledgements} We would like to thank the anonymous referees for their very interesting criticisms and helpful comments about this work. We also thank the SNCF company and especially M. Antoni from the Infrastructure Department for availability of data.

\end{document}